\documentclass[apj]{emulateapj}

\newcommand{\vect}[1]{{\mathbf{#1}}}
\newcommand{\mydotfill}{\leaders\hbox to 2pt{\hss.\hss}\hfill\phantom{.}}

\slugcomment{draft \today}

\shorttitle{Fractal Density Distribution from Solenoidal versus Compressive Forcing}
\shortauthors{Federrath, Klessen, \& Schmidt}

\begin{document}

\title{THE FRACTAL DENSITY STRUCTURE IN SUPERSONIC ISOTHERMAL TURBULENCE: SOLENOIDAL VERSUS COMPRESSIVE ENERGY INJECTION}
\author{Christoph Federrath\altaffilmark{1,2}, Ralf S.~Klessen\altaffilmark{1}, and Wolfram Schmidt\altaffilmark{3}}

\email{chfeder@ita.uni-heidelberg.de}
\email{rklessen@ita.uni-heidelberg.de}
\email{schmidt@astro.uni-wuerzburg.de}

\altaffiltext{1}{Zentrum f\"ur Astronomie der Universit\"at Heidelberg, \\Institut f\"ur Theoretische Astrophysik, Albert-Ueberle-Str.~2, D-69120 Heidelberg, Germany}
\altaffiltext{2}{Max-Planck-Institute for Astronomy, K\"onigstuhl 17, D-69117 Heidelberg, Germany}
\altaffiltext{3}{Lehrstuhl f\"ur Astronomie der Universit\"at W\"urzburg, \\Institut f\"ur Theoretische Physik und Astrophysik, Am Hubland, D-97074 W\"urzburg, Germany}

\begin{abstract}
In a systematic study, we compare the density statistics in high-resolution numerical experiments of supersonic isothermal turbulence, driven by the usually adopted solenoidal (divergence-free) forcing and by compressive (curl-free) forcing. We find that for the same rms Mach number, compressive forcing produces much stronger density enhancements and larger voids compared to solenoidal forcing. Consequently, the Fourier spectra of density fluctuations are significantly steeper. This result is confirmed using the $\Delta$-variance analysis, which yields power-law exponents $\beta\!\sim\!3.4$ for compressive forcing and $\beta\!\sim\!2.8$ for solenoidal forcing. We obtain fractal dimension estimates from the density spectra and $\Delta$-variance scaling, and by using the box counting, mass size and perimeter area methods applied to the volumetric data, projections and slices of our turbulent density fields. Our results suggest that compressive forcing yields fractal dimensions significantly smaller compared to solenoidal forcing. However, the actual values depend sensitively on the adopted method, with the most reliable estimates based on the $\Delta$-variance, or equivalently, on Fourier spectra. Using these methods, we obtain $D\!\sim\!2.3$ for compressive and $D\!\sim\!2.6$ for solenoidal forcing, which is within the range of fractal dimension estimates inferred from observations ($D\!\sim\!2.0\dots2.7$). The velocity dispersion to size relations for both solenoidal and compressive forcings obtained from velocity spectra follow a power law with exponents in the range $0.4\dots0.5$, in good agreement with previous studies.
\end{abstract}


\section{INTRODUCTION}

Observations provide velocity dispersion to size relations for various molecular clouds (MCs), which document the existence of supersonic random motions on scales larger than $\sim\!0.1\,\mathrm{pc}$ \citep[e.g.,][]{Larson1981,Myers1983,PeraultFalgaronePuget1986,SolomonEtAl1987,FalgaronePugetPerault1992,HeyerBrunt2004}. These motions are associated with compressible turbulence \citep[e.g.,][]{ElmegreenScalo2004,ScaloElmegreen2004,MacLowKlessen2004} in the interstellar medium \citep{Ferriere2001} and exhibit a single turbulent cascade or spatially separated coexisting inertial ranges \citep{PassotPouquetWoodward1988} similar to the kinetic energy cascade of incompressible \citet{Kolmogorov1941c} turbulence. However, there are various physical processes (e.g., self-gravity, magnetic fields, nonequilibrium chemistry) and especially the compressibility of the gas, that alter the scaling laws \citep[e.g.,][]{Fleck1996} and statistics \citep[e.g., intermittency corrections measured by][]{HilyBlantFalgaronePety2008} established for incompressible turbulence.

The physical origin and characteristics of the turbulent fluctuations are still a matter of debate. To advance on the question of how turbulence isdriven in the interstellar medium, we present results of high-resolution numerical experiments of supersonic isothermal turbulence comparing two distinct and extreme ways of driving the turbulence in a systematic study: 1) solenoidal forcing (divergence-free or rotational forcing), and 2) compressive forcing (curl-free or dilatational forcing).

Various numerical and analytical studies have provided important insight into the statistics of supersonic isothermal turbulence \citep[e.g.,][]{PorterPouquetWoodward1992,Vazquez1994,PadoanNordlundJones1997,PassotVazquez1998,StoneOstrikerGammie1998,MacLow1999,Klessen2000,OstrikerStoneGammie2001,BoldyrevNordlundPadoan2002,LiKlessenMacLow2003,PadoanJimenezNordlundBoldyrev2004,JappsenEtAl2005,BallesterosEtAl2006,KritsukEtAl2007,LemasterStone2008}. Most of these studies use purely solenoidal or weakly compressive kinetic energy injection mechanisms (forcing) to excite turbulent motions. In the present study, we aim at comparing the usual case of solenoidal (divergence-free) forcing with the case of fully compressive (curl-free) forcing. The actual way of turbulence production in real MCs is expected to be far more complex compared to what we can model with the present simulations, probably consisting of a convolution of various agents producing turbulence, and mixtures of solenoidal and compressive modes \citep[e.g.,][]{ElmegreenScalo2004,MacLowKlessen2004}. Here, we systematically investigate the extreme cases of purely solenoidal versus purely compressive energy injection.

Analyzing the density correlation statistics and fractal structure obtained in our hydrodynamic simulations, we show that compressive forcing leads to significantly steeper density fluctuation spectra and consequently to fractal dimensions of the turbulent gas structures, that are significantly smaller compared to the usually adopted solenoidal forcing. We use Fourier analysis, $\Delta$-variance analysis, structure functions, the fractal mass size, box counting, and perimeter area methods to obtain fractal dimension estimates. We apply the $\Delta$-variance analysis to both our 3-dimensional data and to 2-dimensional projections, and the perimeter area method to projections and slices through the turbulent density structures supporting the result of a significantly smaller fractal dimension for compressive forcing compared to solenoidal forcing. Although compressive forcing yields significantly smaller fractal dimensions than solenoidal forcing, the estimated fractal dimensions are in the range $2.0\dots2.7$ consistent with observational estimates \citep[e.g.,][]{ElmegreenFalgarone1996,SanchezEtAl2007}

We explain our numerical method, construction of solenoidal and compressive forcing fields and fractal analysis techniques in Section~\ref{sec:methods}. In Section~\ref{sec:results}, we show that our results are consistent with previous studies using solenoidal forcing, whereas compressive forcing yields much stronger density contrasts and consequently leads to significantly smaller fractal dimensions. In Section~\ref{sec:conclusions}, we summarize our conclusions.

\section{SIMULATIONS AND METHODS} \label{sec:methods}

The piecewise parabolic method \citep{ColellaWoodward1984} implementation of the astrophysical code FLASH3 \citep{FryxellEtAl2000,DubeyEtAl2008} was used to integrate the hydrodynamic equations on periodic uniform grids with $256^3$, $512^3$ and $1024^3$ grid points. Density $\rho$, velocity $\vect{v}$ and pressure $P$ are related through the equations
\begin{eqnarray}
\frac{\partial \rho}{\partial t} + \nabla \cdot(\rho \vect{v}) & = & 0 \label{eq:hydro1} \\
\frac{\partial \vect{v}}{\partial t} + (\vect{v} \cdot \nabla) \vect{v} & = & -\frac{1}{\rho}\nabla P + \vect{f}\;. \label{eq:hydro2}
\end{eqnarray}
Note that an energy equation is not needed, because we model isothermal gas. The pressure is simply given by $P=c_s^2\rho$ with the constant sound speed $c_s$. Isothermality is a very crude, but reasonable first approximation for modeling the thermodynamic behavior of MCs \citep{WolfireEtAl1995,PavlovskiSmithMacLow2006}. Due to the isothermal approximation, the hydrodynamic equations are scale-free, and we can solve them for a chosen density scale $\rho_0=1$ (mean density), sound speed $c_s=1$ and domain size $L=1$. The only remaining free parameter therefore is the dimensionless rms Mach number $\mathcal{M}$, which can be varied. It is important to note that the forcing term $\vect{f}$ used to drive turbulent motions appearing as source term in equation~(\ref{eq:hydro2}) can also be varied. In the present study, we vary the forcing term, investigating the difference between purely solenoidal and purely compressive kinetic energy injection, while keeping the rms Mach number fixed.

Equations~(\ref{eq:hydro1}) and~(\ref{eq:hydro2}) have been solved numerically with periodic boundary conditions using an isothermal equation of state in the context of MC dynamics in various studies \citep[e.g.,][]{PadoanNordlundJones1997,PassotVazquez1998,StoneOstrikerGammie1998,MacLowEtAl1998,MacLow1999,KlessenHeitschMacLow2000,HeitschMacLowKlessen2001,BoldyrevNordlundPadoan2002,LiKlessenMacLow2003,PadoanJimenezNordlundBoldyrev2004,JappsenEtAl2005,BallesterosEtAl2006,KritsukEtAl2007,DibEtAl2008,OffnerKleinMcKee2008}. We aim at comparing turbulence statistics obtained in our study with the results of these studies. In particular, we want to check the influence of different forcings. Therefore, we concentrate on two extreme cases: 1) the usually adopted solenoidal forcing (divergence-free forcing) and 2) fully compressive forcing (curl-free forcing).

\subsection{Forcing Module}
Turbulent fluctuations have to be excited and maintained in order to study stationary turbulence statistics in detail. If not constantly driven by a random force field, turbulent motions damp due to dissipation. In most studies, the force field is constructed in Fourier space by a 3-dimensional stochastic procedure \citep[e.g.,][]{DubinskiNarayanPhillips1995,MacLowEtAl1998,StoneOstrikerGammie1998}, which generates a random vector field $\vect{f}$ after Fourier transformation back into physical space. This field will on average contain $2/3$ of its energy in solenoidal modes (transversal modes) and $1/3$ in compressive modes (longitudinal modes), because in 3-dimensional space, waves have two spatial directions for the transversal part, whereas the longitudinal part has only one \citep[see, e.g.,][]{ElmegreenScalo2004}. In order to obtain a purely solenoidal, or a purely compressive forcing field $\vect{f}$, a Helmholtz decomposition can be made by applying the projection operator $\mathcal{P}_{ij}^\zeta$ in Fourier space ($k$-space)
\begin{equation} \label{eq:projectionoperator}
\mathcal{P}_{ij}^\zeta=\zeta\mathcal{P}_{ij}^\perp+(1-\zeta)\mathcal{P}_{ij}^\parallel=\zeta\delta_{ij}+(1-2\zeta)\frac{k_i k_j}{|k|^2}
\end{equation}
prior to the inverse Fourier transformation into real space. By setting the parameter $\zeta\in[0,1]$ one can adjust the mixture of solenoidal and compressive modes. If we set $\zeta=1$, $\mathcal{P}_{ij}^\zeta$ projects only the solenoidal component, whereas only the compressive component is obtained by setting $\zeta=0$.

The forcing term $\vect{f}$ is typically either modeled as a spatially static pattern with time-dependent amplitude \citep[following the recipes, e.g., by][]{MacLowEtAl1998,StoneOstrikerGammie1998} or by using an Ornstein-Uhlenbeck (OU) process \citep[e.g.,][]{EswaranPope1988,SchmidtEtAl2006}, which modulates the pattern smoothly in space and time on a well-defined autocorrelation timescale $T$ resulting in a constant energy input rate. We follow the usual approach and set the autocorrelation timescale equal to the dynamical timescale $T=L/(2V)$, where $L$ is the size of the computational domain, $V=c_s\mathcal{M}$ and $\mathcal{M}\approx5.5$ is the rms Mach number in all runs. Therefore, $T$ is the time for the most energetic fluctuations (at $k=2$ in Fourier space, which corresponds to $L/2$) to cross half of the box. It is furthermore equal to the decay time constant of the turbulence \citep{StoneOstrikerGammie1998,MacLow1999}. The forcing amplitude follows a parabolic power spectrum only containing power on the largest scales in a small interval of wavenumbers $1<k<3$ peaking at $k=2$. The influence of varying the scale of energy input has been investigated for instance by \citet{MacLow1999}, \citet{KlessenHeitschMacLow2000}, \citet{HeitschMacLowKlessen2001} and \citet{VazquezBallesterosKlessen2003}. Here, we only consider the usually applied large-scale stochastic forcing. This way of forcing models the kinetic energy input from larger scale turbulent fluctuations breaking up into smaller structures and feeding kinetic energy to smaller scales.

We checked that our results are not sensitive to the particular method for generating turbulent motions, i.e., by using an almost static pattern (using a very large autocorrelation time in the OU process), and by using a band spectrum instead of a parabolic Fourier spectrum for forcing. Variations in the spectral form of the large scale forcing did not significantly change the results obtained in the present study. However, changing the mixture of modes from a purely solenoidal to a purely compressive forcing always yielded significant differences.

\subsection{Initial Conditions and Post Processing}
Starting from a uniform density distribution and zero velocity, the forcing excites turbulent motions. The forcing amplitude is adjusted to excite turbulence with rms Mach number $\mathcal{M}\!\sim\!5.5$. We use $\mathcal{M}$ as the control parameter, because this dimensionless number is often expected to solely determine physical properties of scale-invariant turbulent flows. The purpose of the present study was to determine the effect of varying the forcing from purely solenoidal to purely compressive, so we keep the rms Mach number fixed besides all other parameters and varied only the forcing between solenoidal and compressive.

Equations~(\ref{eq:hydro1}) and~(\ref{eq:hydro2}) were evolved for ten dynamical timescales $T$, which allows us to study a large sample of statistically stationary realizations of the turbulent flow. We wait for two dynamical timescales before averaging all statistical measures in the time interval $2\leq t/T \leq 10$. Since we have produced snapshots every $0.1\,T$, the resulting statistical sample consists of $81$ realizations of the turbulent field. The averaging procedure is important to derive meaningful statistics, because all quantities are subject to statistical fluctuations \citep[e.g.,][]{KritsukEtAl2007}. The averaging procedure furthermore provides a handle on the 1$\sigma$ temporal fluctuations between different realizations. Unless otherwise stated, the 1$\sigma$ temporal fluctuations are indicated as error bars in the results section.

\subsection{Box Counting Method} \label{sec:box-counting}
We analyzed fractal structures in our simulation data using the box counting method. In the first step, the fractal structure is defined by marking all cells belonging to the fractal set, if they are above a certain density threshold, whereas all cells below that threshold are marked as not belonging to the fractal structure. In the second step, the structure as defined above was scanned by applying a box (mask) of size $l$ and counting how often the structure is covered by that box. This procedure was repeated varying the size of the box resulting in a set of counts $N_i$ and box sizes $l_i$. A plot of $\log(N_i)$ against $\log(l_i)$ often reveals a scaling range over which the points fall close to a straight line with the box counting dimension $D_b$ as the negative slope of that line \citep[e.g.,][]{MandelbrotFrame2002,PeitgenEtAl2004}. This implies a power-law scaling $N(l)\propto l^{-D_b}$ in the scaling range.

Setting the density threshold $\rho_\mathrm{th}$ for defining the fractal structure is obviously a critical choice. Using $\rho_\mathrm{th}=0$ naturally results in $D_b=3$, whereas setting $\rho_\mathrm{th}=\rho_\mathrm{max}$ leads to $D_b=0$. We computed the box counting dimension for different density thresholds and discuss its influence on the results.

\subsection{Mass Size Method} \label{sec:mass-size}
The fractal mass size dimension was obtained by computing the mass contained inside concentric boxes with increasing box size $l$ centered on the densest cells of the data set and averaging over cells with $\rho>\rho_\mathrm{max}/2$, following the method described by \citet{KritsukEtAl2007}. This yields a set of masses $M_i$ and box sizes $l_i$. A plot of $\log(M_i)$ against $\log(l_i)$ often reveals a scaling range over which the points fall close to a straight line with the mass size dimension $D_m$ as the slope of that line \citep[e.g.,][]{MandelbrotFrame2002,PeitgenEtAl2004}. This implies a power-law scaling $M(l)\propto l^{D_m}$ in the scaling range.

Power-law relations of the form $M(l)\propto l^a$ should be considered with caution in the context of fractals, because such relations often occur in physics and do not necessarily imply that $a$ is a fractal dimension, i.e., for a 3-dimensional density distribution, $a>3$ can occur \citep[see][]{Mandelbrot1983,ElmegreenFalgarone1996}.

\subsection{Perimeter Area Method} \label{sec:perimeter-area}
We also applied the perimeter area method, which is frequently applied for measuring the fractal dimension of interstellar gas clouds. The boundary curves and areas of coherent structures with equal density were identified in both, 2-dimensional projections and 2-dimensional slices through the computational domain. Varying the density threshold yields a set of structures with perimeters $\mathcal{P}_i$ and areas $\mathcal{A}_i$. Fitting a power law of the form $\mathcal{P}\propto\mathcal{A}^{D_p/2}$ (log-log plot as for the box counting and mass size dimensions) yields the perimeter area dimension $D_p$. Structures with very smooth boundary curves exhibit $D_p=1$, whereas for structures with totally convoluted perimeters, $\mathcal{P}$ grows linearly with the area occupied by the structure resulting in $D_p=2$.

\section{RESULTS AND DISCUSSION} \label{sec:results}

\subsection{Time Evolution}
\begin{figure*}[t]
\begin{center}
\begin{tabular}{rr}
\includegraphics[width=0.48\linewidth]{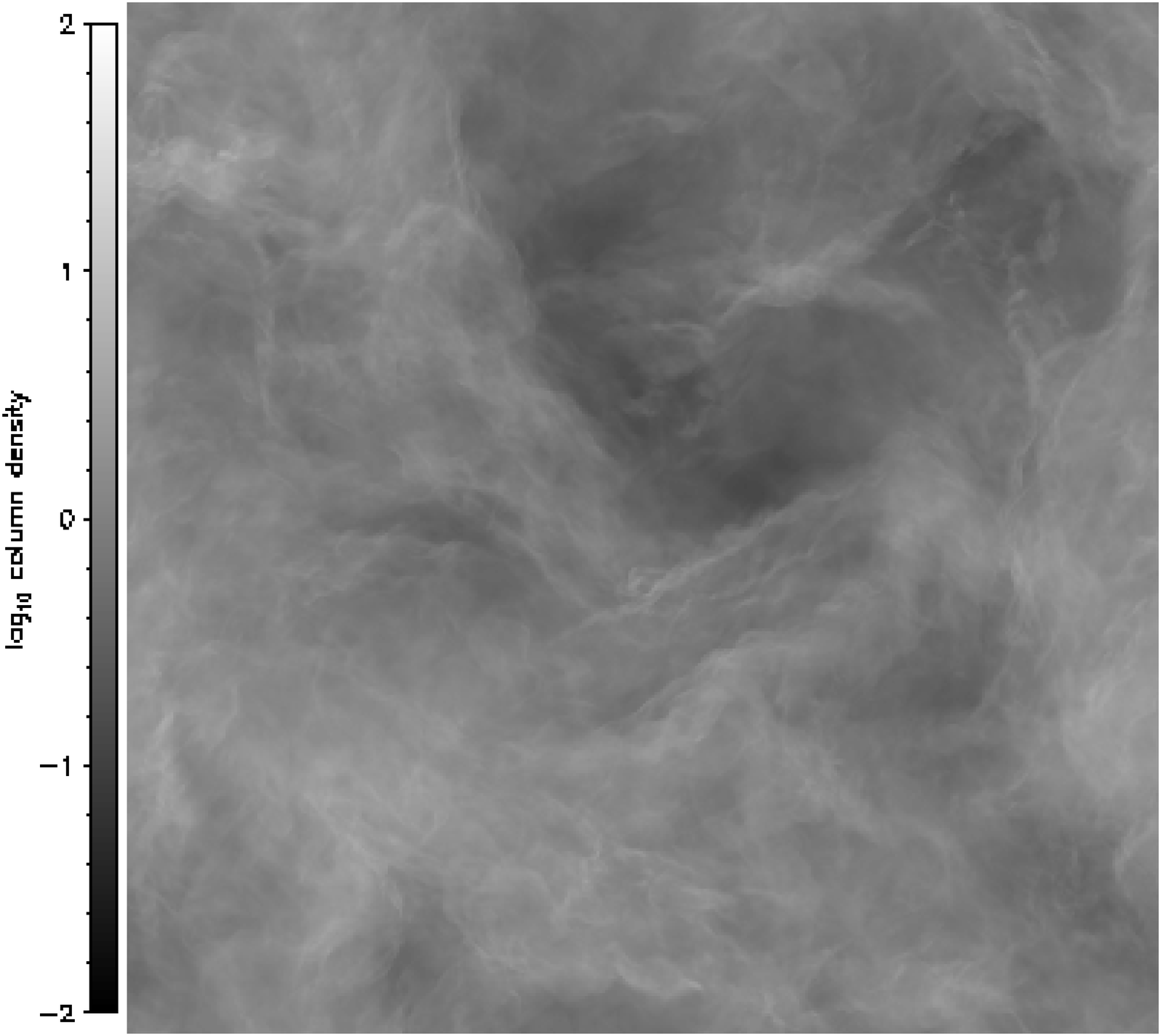} &
\includegraphics[width=0.48\linewidth]{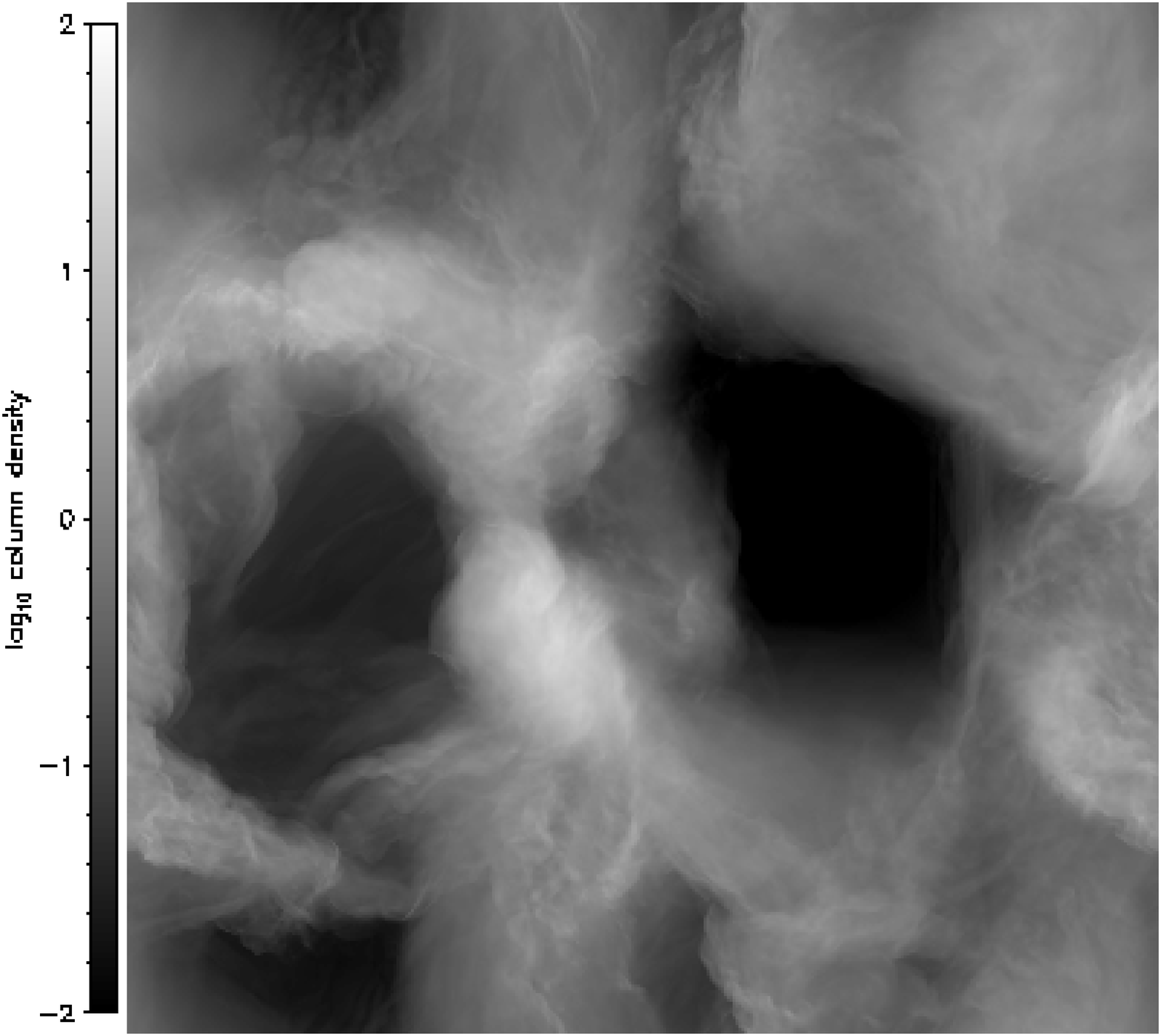} \\
\includegraphics[width=0.48\linewidth]{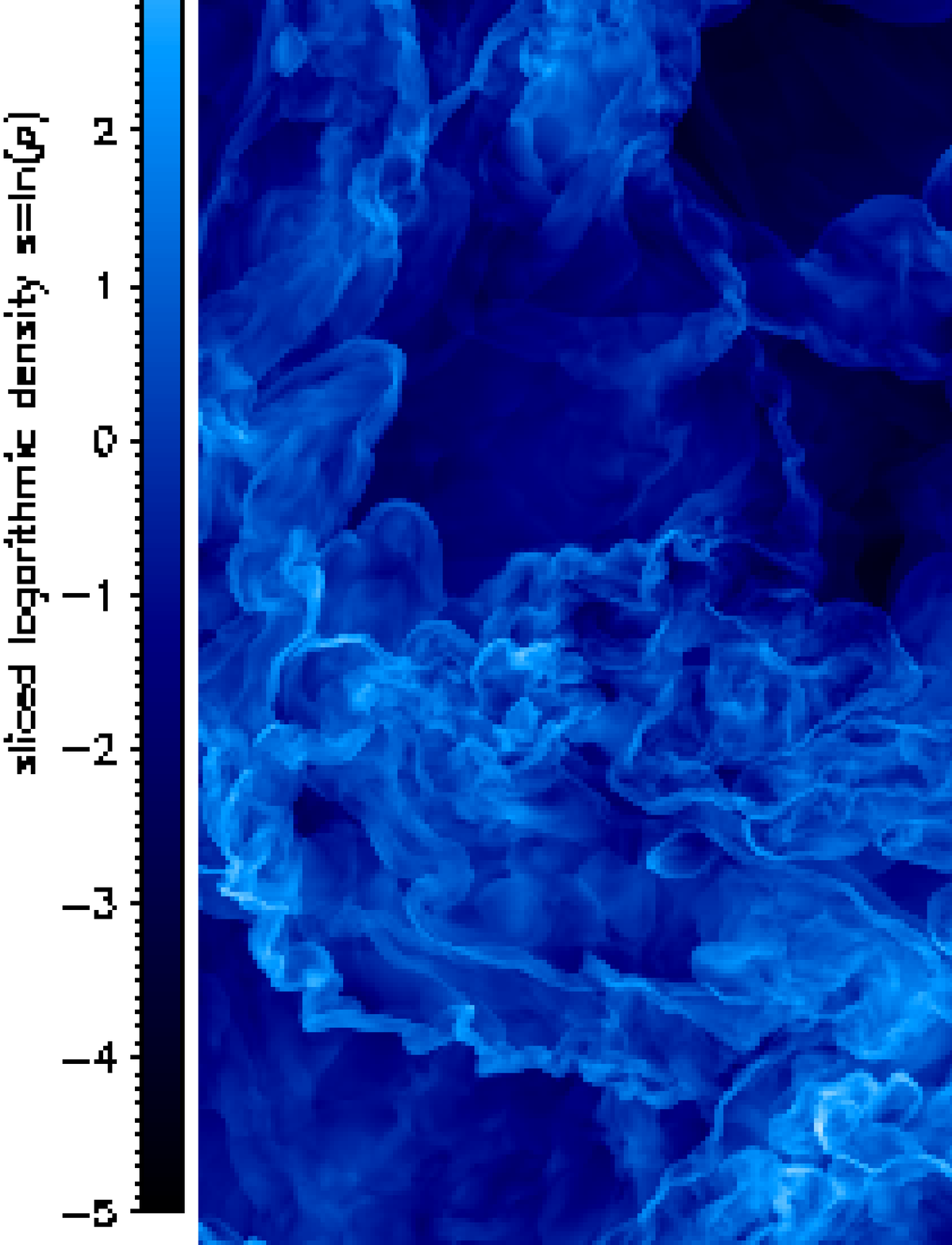} &
\includegraphics[width=0.48\linewidth]{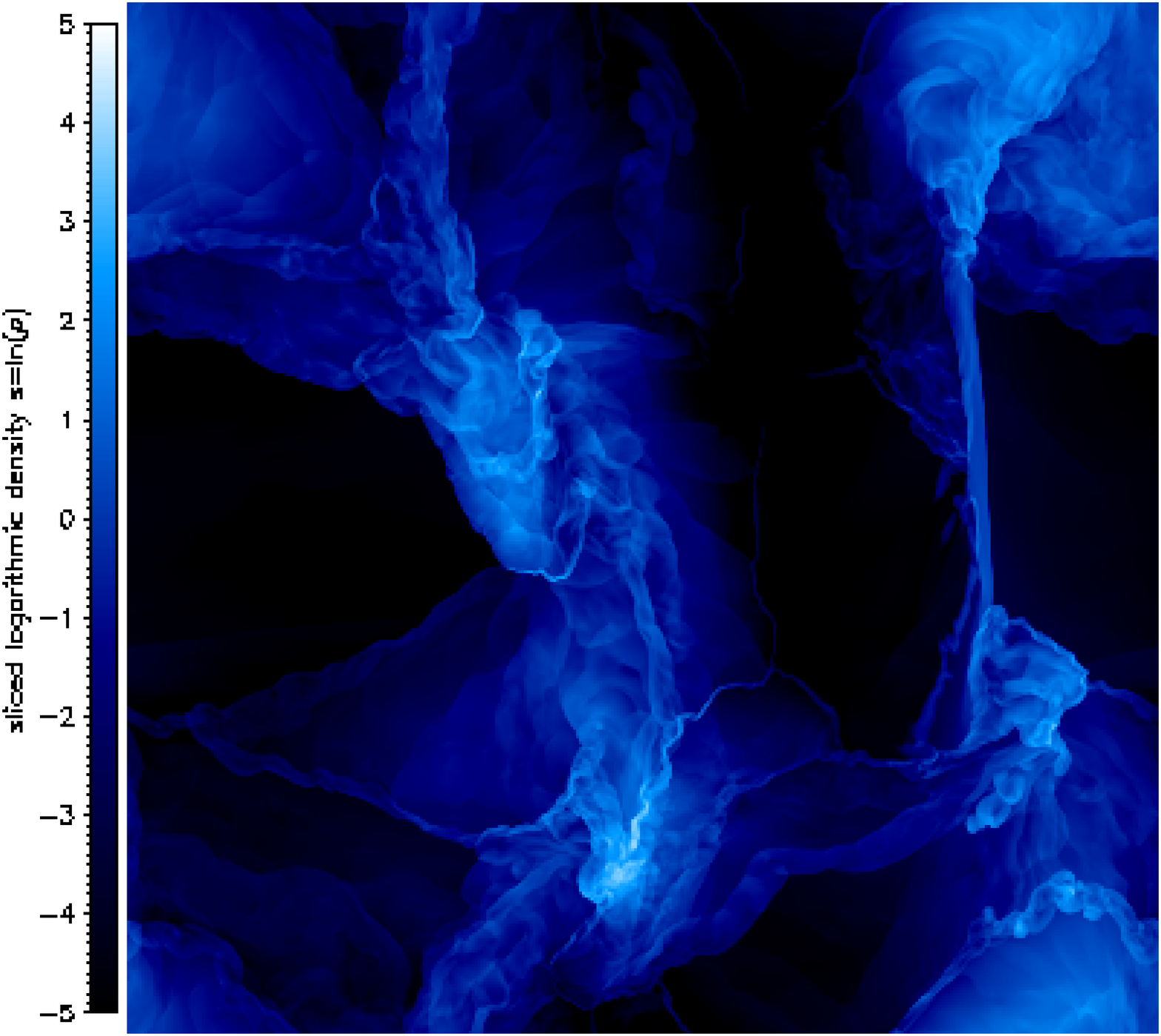}
\end{tabular}
\end{center}
\caption{\emph{Top panels:} Column density computed along the $z$-axis in units of the mean column density for solenoidal forcing (\emph{left}), and compressive forcing (\emph{right}) at a randomly picked time $t=5\,T$ in the regime of statistically stationary compressible turbulence. Both maps show 4 orders of magnitude in column density with the same scaling for direct comparison of solenoidal and compressive forcing at rms Mach number $\sim\!5.5$. \emph{Bottom panels:} Same as top panels, but slices through the density field at $z=0$. Compressive forcing yields stronger density enhancements and larger voids compared to solenoidal forcing.}
\label{fig:snapshots}
\end{figure*}

\begin{figure}[t]
\centerline{\includegraphics[width=0.7\linewidth]{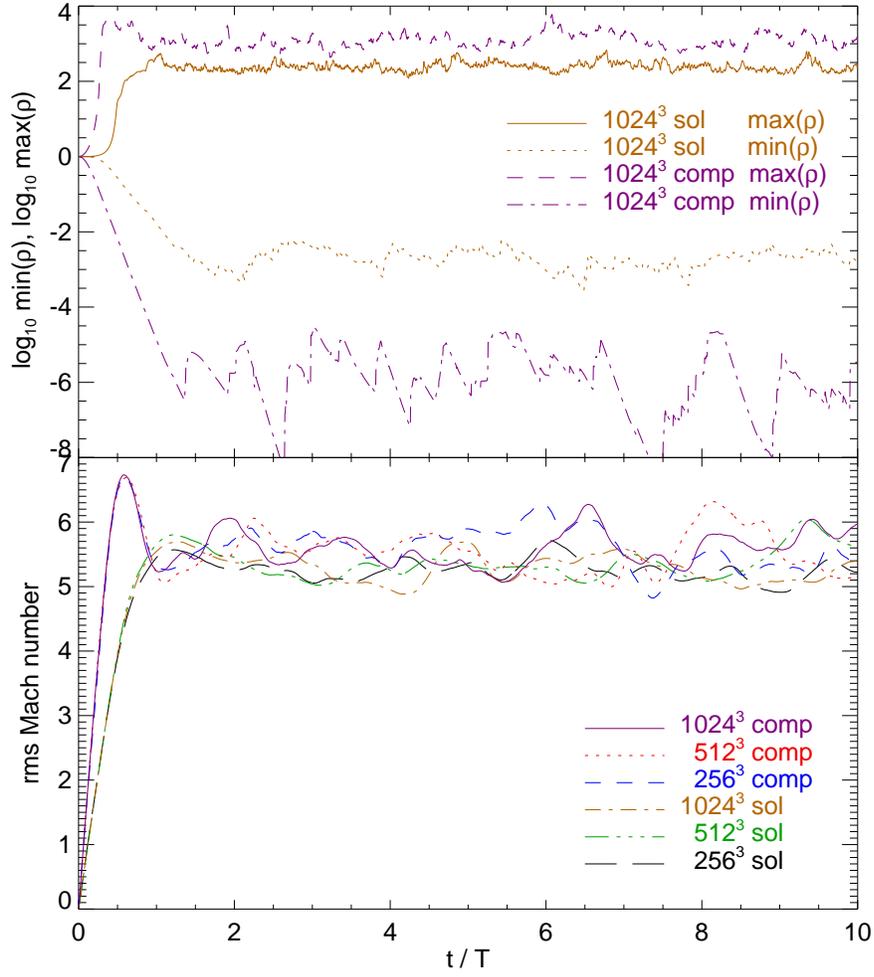}}
\caption{\emph{Bottom}: rms Mach number $\mathcal{M}$ as function of the dynamical time $T$ for $256^3$, $512^3$ and $1024^3$ numerical grid resolution. \emph{Top}: Minimum and maximum density as function of the dynamical time $T$. At $\sim\!2\,T$, a statistically stationary state was reached for both solenoidal (sol) and compressive (comp) forcing. Consequently, we can use all the available 81 snapshots within $2 \leq t/T \leq 10$ for averaging statistical measures (e.g., Fourier spectra, structure functions, $\Delta$-variance, fractal perimeter area, box counting and mass size analysis) to improve statistical significance and to compute corresponding 1$\sigma$ temporal fluctuations. Note that on average, the maximum density is almost $\sim\!10$ times larger for compressive forcing compared to solenoidal forcing, although the rms Mach number is roughly the same in both cases. The maximum density is subject to strongly intermittent fluctuations \citep[e.g.,][]{KritsukEtAl2007} for both solenoidal and compressive forcing.}
\label{fig:timeevol}
\end{figure}

Figure~\ref{fig:snapshots} compares projections (top panels) and slices (bottom panels) of the density field in the $x$-$y$-plane from a randomly picked snapshot ($t\,=\,5\,T$) for solenoidal versus compressive forcing as an example of the typical density structure in the state of statistically stationary supersonic turbulent flow. This regime was safely reached after $2$ dynamical times $T$, which is demonstrated in Figure~\ref{fig:timeevol}. The rms Mach number has settled to $\mathcal{M}\!\sim\!5.5$ for both solenoidal and compressive forcing, and for numerical resolutions of $256^3$, $512^3$ and $1024^3$ grid points after $2$ dynamical times. Not only the velocity statistics has converged to a stationary state, but also the density statistics, which is shown in terms of minimum and maximum densities in the top panel of Figure~\ref{fig:timeevol}. Obviously, compressive forcing produces larger density contrasts, which results in higher density peaks and larger voids. Although both cases exhibit the same rms Mach number, the solenoidally driven case gives a much smoother density distribution with smaller dispersion. In both cases, the maximum density is subject to strong intermittent fluctuations \citep[e.g.,][]{FalgaroneEtAl1994,KritsukEtAl2007} leading to temporal variations in the maximum density of order one magnitude.

\subsection{Fourier Spectrum Functions} \label{sec:fourierspectra}

\begin{figure}[t]
\centerline{\includegraphics[width=0.7\linewidth]{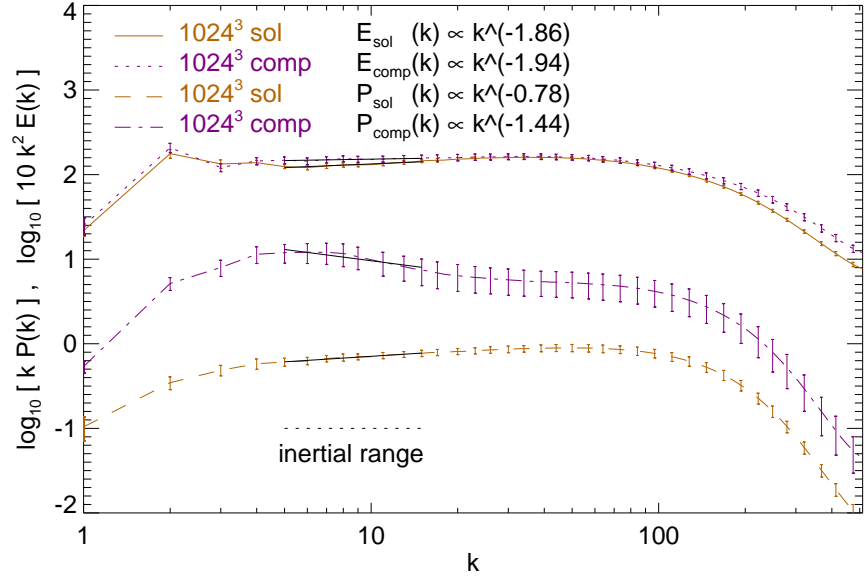}}
\caption{Kinetic energy Fourier spectra $E(k)$ compensated by $k^2$ corresponding to Burgers turbulence (upper curves), and density fluctuation Fourier spectra compensated by $k$ (lower curves) for solenoidal and compressive forcing respectively. Power-law fits in the inertial range $5\lesssim k \lesssim 15$ are shown as thin solid lines. The velocity power spectra exhibit only marginal differences between solenoidal and compressive forcing. The scaling of the density power spectra on the other hand differs significantly comparing both forcings. Accordingly, the stepper density power spectrum for compressive forcing leads to a smaller fractal box coverage dimension $D_b\!\sim\!2.28$ compared to $D_b\!\sim\!2.61$ for solenoidal forcing.}
\label{fig:fourierspectra}
\end{figure}

We begin the analysis of spatial correlations by showing 1-dimensional Fourier spectra derived for solenoidal and compressive forcing. Figure~\ref{fig:fourierspectra} presents a comparison of velocity Fourier spectra
\begin{equation}
E(k)\,dk = \frac{1}{2}\int\widehat{\vect{v}}\cdot\widehat{\vect{v}}^{*}\,4\pi k^2 dk
\end{equation}
and density fluctuation Fourier spectra
\begin{equation}
P(k)\,dk = \int\widehat{(\rho-\rho_0)}\widehat{(\rho-\rho_0)}^{*}\,4\pi k^2 dk\,.
\end{equation}
These were computed with the standard method \citep[e.g.,][]{Frisch1995}, i.e., by integration over spherical shells in Fourier space using the Fourier transformed velocity components $\widehat{v_i}(k)$ and the Fourier transformed density fluctuations $\widehat{\rho-\rho_0}(k)$, where $\rho_0$ denotes the mean density. Velocity Fourier spectra $E(k)$ are typically used to distinguish between \citet{Kolmogorov1941c} turbulence, $E(k)\propto k^{-5/3}$ and Burgers turbulence, $E(k)\propto k^{-2}$ in the inertial range. For highly compressible isothermal supersonic turbulent flow, it has been shown that the inertial range scaling is closer to Burgers turbulence. For instance, \citet{KritsukEtAl2007} find $E(k)\propto k^{-1.95}$ and \citet{SchmidtEtAl2008} measured $E(k)\propto k^{-1.87}$. In the present study, we obtain $E(k)\propto k^{-1.94}$ for compressive forcing, slightly steeper than $E(k)\propto k^{-1.86}$ for solenoidal forcing. Taking the error bars (temporal variations) into account, this represents just marginal difference between both forcing schemes, and our estimates within the inertial range are consistent with \citet{KritsukEtAl2007} and \citet{SchmidtEtAl2008}. Table~1 summarizes all results obtained for solenoidal and compressive forcing providing the formal least-squares fitting errors, which were obtained by taking into account the 1$\sigma$ temporal fluctuations. Table~1 furthermore provides a summary of defining relations for the following fractal dimension estimates.

Note that similar to \citet{KritsukEtAl2007} and \citet{SchmidtEtAl2008}, we define our inertial range in a very small range of wavenumbers $5 \lesssim k \lesssim 15$ because even at numerical resolutions of $1024^3$ grid points, the inertial range is very small \citep[see, e.g.,][]{KleinEtAl2007} due to the bottleneck effect \citep[e.g.,][]{DoblerEtAl2003,HaugenBrandenburg2004,SchmidtEtAl2006,KritsukEtAl2007}, which affects the Fourier spectrum in the dissipation range. For our simulations, we estimate that wavenumbers $k\gtrsim 20$ may be affected by the bottleneck. The influence of the bottleneck effect is less pronounced in physical space, which allows for a somewhat larger fitting range for scaling estimates obtained in physical space \citep[similar to, e.g.,][]{KowalLazarian2007,KritsukEtAl2007,SchmidtEtAl2008} for the $\Delta$-variance analysis in Section~\ref{sec:deltavar}.

The density fluctuation spectra in Figure~\ref{fig:fourierspectra} show considerable differences in their amplitude and inertial range scaling for solenoidal and compressive forcing. First, we discuss the difference in the amplitudes. As discussed in \citet{FederrathKlessenSchmidt2008}, the standard deviation of the density probability distribution function (PDF) is $\sim\!$ three times larger for compressive forcing compared to solenoidal forcing. This result is recovered in the present analysis by integrating the density fluctuation spectra
\begin{equation}
\sigma_\rho^2 = \sum_{i=1}^n (\rho_i-\rho_0)^2 = \int P(k)\,dk\,,
\end{equation}
where $n=1024^3$ is the number of grid points, which yields the standard deviation $\sigma_\rho$. For solenoidal forcing, we compute $\int P(k)dk\!\sim\!1.89$, whereas for compressive forcing, $\int P(k)dk\!\sim\!5.93$, in very good agreement with the standard deviations $\sigma_\rho\!\sim\!1.89$ and $\sigma_\rho\!\sim\!5.86$ obtained from the PDFs in \citet{FederrathKlessenSchmidt2008}.

Second, the inertial range scaling differs significantly for the two cases. For solenoidal forcing, $P(k)\propto k^{-0.78}$ and for compressive forcing, $P(k)\propto k^{-1.44}$. Our inertial range scaling inferred for solenoidal forcing is consistent with the weakly magnetized super-Alfv\'enic supersonic MHD models from \citet[][Tab.~2]{KowalLazarianBeresnyak2007} using solenoidal forcing. \citet{KowalLazarianBeresnyak2007} find $P(k)\propto k^{-0.6\pm0.2}$ for their model B.1P.01 with very weak magnetic field, slightly larger rms Mach number ($\mathcal{M}\!\sim\!7.1$), and resolution\footnote{The resolution dependence for our simulations is discussed in Section~\ref{sec:deltavar}.} of $256^3$ explaining the small differences comparing their result with ours. It is furthermore in agreement with the purely hydrodynamic estimates by \citet{KritsukNormanPadoan2006} with resolutions up to $2048^3$ using adaptive mesh refinement, who obtain $P(k)\propto k^{-0.8\dots0.9}$.

Note that in general, the power-law exponents $\alpha$ of 1-dimensional Fourier spectra are related to the power-law exponents $\beta$ of the corresponding 3-dimensional Fourier power spectra through $\beta=\alpha+2$. As discussed by \citet{StutzkiEtAl1998}, the power-law scaling of the density spectrum is furthermore related to the fractal drift exponent $H$ (Hurst exponent). Considering a 1-dimensional density power spectrum $P(k)\propto k^{-\alpha}$ leads to a Hurst exponent of $H=\alpha/2=(\beta-2)/2$. The Hurst exponent is related to the fractal box coverage dimension $D_b=E+1-H$ with the dimensionality $E=2$ for the box coverage of a fractal surface embedded in 3-dimensional space \citep{StutzkiEtAl1998}. Consequently, we obtain $H\!\sim\!0.39$ and $D_b\!\sim\!2.61$ for solenoidal forcing, and $H\!\sim\!0.72$ and $D_b\!\sim\!2.28$ for compressive forcing. Thus, the fractal Hurst exponent is significantly smaller for compressive forcing. The estimates for the Hurst exponents and the corresponding relations with box counting and perimeter area dimensions are summarized in Table~1.

\subsection{$\Delta$-Variance Analysis} \label{sec:deltavar}

\begin{figure}[t]
\centerline{\includegraphics[width=0.7\linewidth]{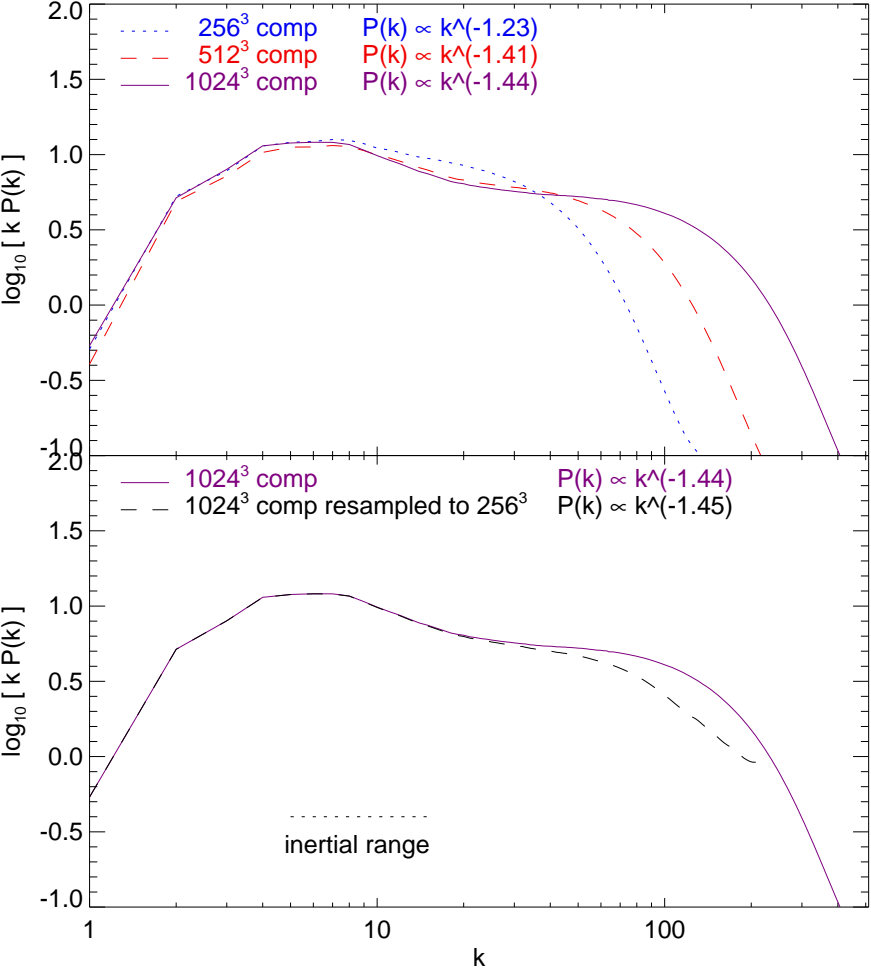}}
\caption{\emph{Top panel:} Numerical resolution comparison of density fluctuation Fourier spectra for compressive forcing. At $512^3$ and $1024^3$, the spectra are almost converged with resolution, whereas the $256^3$ run deviates by $\sim\!15\%$. \emph{Bottom panel}: Density fluctuation Fourier spectra at $1024^3$ in comparison with its resampled version to $256^3$ cells. The resampling had virtually no influence on our results within the inertial range.}
\label{fig:resolutionresampling}
\end{figure}

In this section, we present results of the $\Delta$-variance method, which provides an independent measure of the scaling of the density Fourier spectra. \citet{StutzkiEtAl1998} introduced a wavelet transformation called $\Delta$-variance for measuring the exponent of the Fourier spectrum. As complementary analysis, we performed the $\Delta$-variance with the tool developed and provided by \citet{OssenkopfKripsStutzki2008a}. It is an improved version of the original $\Delta$-variance \citep{StutzkiEtAl1998,BenschStutzkiOssenkopf2001}. The $\Delta$-variance measures the amount of structure on a given length scale $l$, filtering the data set $\Phi(\vect{x})$ by applying an up-down-function $\bigodot\!_l$ (typically a French-hat or Mexican-hat filter) of size $l$ and computing the variance of the filtered data set. The $\Delta$-variance is defined as
\begin{equation}
\sigma_\Delta^2(l) = \left<\left(\Phi(\vect{x})\ast\bigodot\!\frac{\!}{\!}_l(\vect{x})\right)^2\right>_{\!\vect{x}}\,,
\end{equation}
where the average is computed over all valid data points at positions $\vect{x}$, and the operator $\ast$ stands for the convolution. The data set can have arbitrary dimensionality. In the present study, we apply the $\Delta$-variance to both, 2-dimensional (projections) and 3-dimensional data sets. We checked the influence of varying the filter function from French-hat to Mexican-hat, as well as varying the diameter ratio of the filter, which yielded no significant differences. The choice of the filter function and the best choice for its diameter ratio is discussed by \citet{OssenkopfKripsStutzki2008a}. Here, we use the original French-hat filter with a diameter ratio of $3.0$ as it has been used before \citep[e.g.,][]{StutzkiEtAl1998,MacLowOssenkopf2000,OssenkopfKlessenHeitsch2001,OssenkopfMacLow2002,OssenkopfEtAl2006}. Note that originally, \citet{StutzkiEtAl1998} applied the $\Delta$-variance to 2-dimensional integrated maps for comparison with observations. Although we have access to the 3-dimensional data from our simulations, we nevertheless computed column density maps and applied the $\Delta$-variance to both the 2-dimensional and 3-dimensional data to determine the effect of projection for applying the $\Delta$-variance. Prior to the 3-dimensional analysis, we resampled the density data cubes with $1024^3$ grid points to a resolution of $256^3$ due to performance issues of the $\Delta$-variance, which is not (yet) a parallelized tool. The resampling to lower resolution is not expected to cause deviations in the inertial range scaling. As long as the simulation itself had enough spatial resolution to resolve the inertial range scaling, the resampling to lower resolution only affects the dissipation range leaving density spectra almost up to the new Nyquist frequency \citep[e.g.,][]{PadoanEtAl2006}. We explicitly show in the bottom panel of Figure~\ref{fig:resolutionresampling} that the resampling indeed did not affect the inertial range. Only the compressive forcing case is shown but the resampling for the solenoidal case exhibits similar behavior.

The upper panel of Figure~\ref{fig:resolutionresampling} shows the influence of varying the numerical resolution of the simulation. The inertial range scaling depends on the numerical resolution. A resolution of $256^3$ grid points seems insufficient to resolve the \emph{exact} inertial range scaling, although the $15\%$ difference compared to the $1024^3$ simulation is of the order of the temporal fluctuations, whereas the difference between solenoidal and compressive forcing (Fig.~\ref{fig:fourierspectra}) is significant. At resolutions of $512^3$ and $1024^3$ grid points, the best-fit power-law scaling agrees quite well, indicating almost numerical convergence. A similar conclusion can be drawn from the density Fourier spectra presented by \citet[][Fig.~4]{KritsukNormanPadoan2006} computed for solenoidal forcing.

\begin{figure}[t]
\centerline{\includegraphics[width=0.7\linewidth]{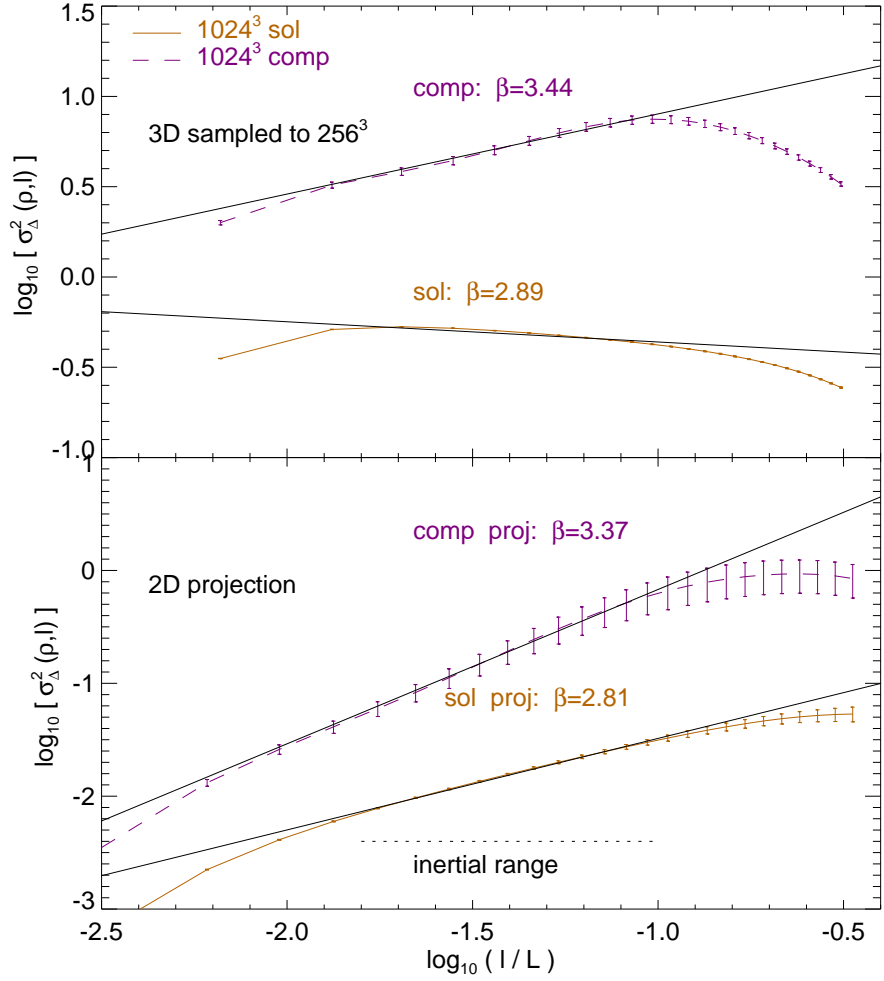}}
\caption{\emph{Top panel:} $\Delta$-variance analysis for the 3-dimensional data set resampled to a resolution of $256^3$ grid cells. \emph{Bottom panel:} $\Delta$-variance applied to the 2-dimensional projections of the $1024^3$ data set. As shown by \citet{StutzkiEtAl1998}, the power-law scaling within the inertial range is preserved upon projection and agrees with the scaling of the Fourier power spectra of Figure~\ref{fig:fourierspectra} within the uncertainties from temporal fluctuations (see Tab.~1).}
\label{fig:deltavar}
\end{figure}

The results of the $\Delta$-variance are presented in Figure~\ref{fig:deltavar}. In the top panel, we show the $\Delta$-variance applied to the 3-dimensional data resampled to $256^3$ grid cells, whereas the bottom panel presents the $\Delta$-variance applied to projections averaged along all three spatial axes. The variation for different projections is almost negligible compared to the temporal fluctuations. We nevertheless added the variation due to projection along the three different axes to the 1$\sigma$ error bars due to temporal fluctuations. Following \citet{StutzkiEtAl1998}, the values of the best-fit power-law exponents $\beta$ are shown corresponding to the 3-dimensional Fourier spectra. Note that the slope $\alpha$ fitted to the $\Delta$-variance is related to the slope of the 3-dimensional Fourier spectrum by $\beta=\alpha+E$, where $E=2$ for the projected data and $E=3$ for the 3-dimensional data resulting in the same power-law exponent $\beta$. As shown by \citet{StutzkiEtAl1998}, the power law scaling of the Fourier spectrum should be preserved at projection to lower dimensions. In agreement with the results by \citet{MacLowOssenkopf2000}, we find that the slopes $\beta\!\sim\!2.89$ (3D) and $\beta\!\sim\!2.81$ (2D projection) for solenoidal forcing, and $\beta\!\sim\!3.44$ (3D) and $\beta\!\sim\!3.37$ (2D projection) for compressive forcing are almost preserved during projection (see Table~1).

We can furthermore check whether the $\Delta$-variance results agree with the Fourier power spectra, since the $\Delta$-variance is supposed to measure the power-law exponent of the Fourier spectrum. As shown in Figure~\ref{fig:fourierspectra}, the density spectra exhibit power laws in the inertial range corresponding to 3-dimensional Fourier power-law exponents $\beta=\alpha+2=2.78$ for solenoidal and $\beta=3.44$ for compressive forcing in consistency with the $\Delta$-variance. Therefore, the $\Delta$-variance confirms the results obtained by the density Fourier spectra, showing that compressive forcing yields a systematically steeper density spectrum compared to solenoidal forcing.

\subsection{Structure Functions}

\begin{figure}[t]
\centerline{\includegraphics[width=0.7\linewidth]{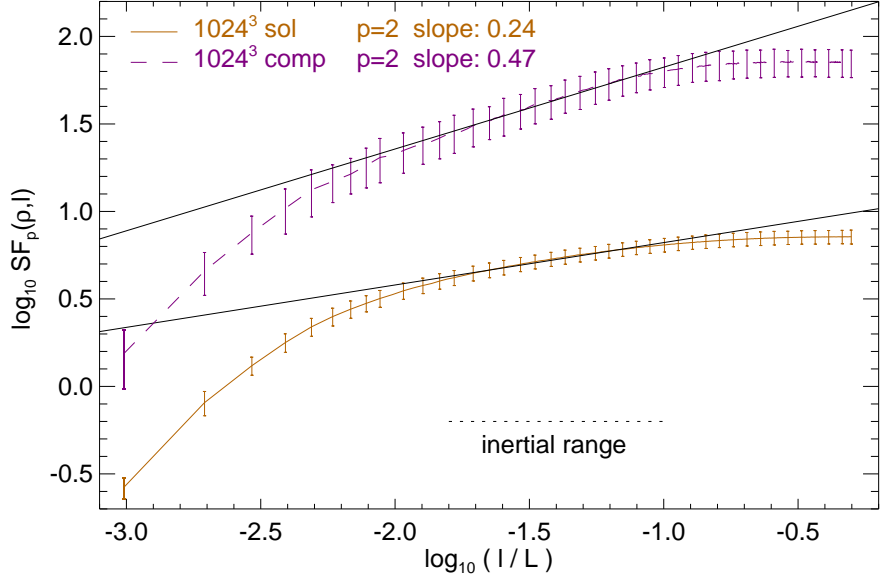}}
\caption{Second order structure functions of the density $\rho$ for solenoidal and compressive forcing. The absolute values of the structure functions are in agreement with the measures of the power spectra and PDFs. The inertial range scaling, however, agrees with the Fourier spectra for compressive forcing only. Solenoidal forcing exhibits a density Fourier spectrum with power-law exponent $\alpha<1$ (Fig.~\ref{fig:fourierspectra}), which precludes the determination of the power-law exponent via structure function analysis \citep[][Appendix~B]{StutzkiEtAl1998}.}
\label{fig:sf}
\end{figure}

Besides the Fourier spectra and the $\Delta$-variance analyzed in the previous sections, structure functions are often used to measure spatial correlations in turbulent velocity and density fields \citep[e.g.,][]{BoldyrevNordlundPadoan2002,PadoanEtAl2003,EsquivelLazarian2005,KritsukEtAl2007,SchmidtEtAl2008,HilyBlantFalgaronePety2008}. Here, we restrict our analysis to the 2nd order structure functions of the density field for comparison with the Fourier spectra and $\Delta$-variances. The 2nd order structure function of the density is defined as
\begin{equation}
\mathrm{SF}_2(l) = \left<\left|\rho(\vect{x})-\rho(\vect{x}+\vect{l})\right|^2\right>_{\!\vect{x}}\,.
\end{equation}
Figure~\ref{fig:sf} presents the 2nd order density structure functions for compressive and solenoidal forcing respectively. In the following, we draw the connection of these structure functions to the power spectra and $\Delta$-variance. One feature of the structure function is its relation to the autocorrelation function $A(l)$ \citep{StutzkiEtAl1998}:
\begin{equation}
\mathrm{SF}_2(l) = 2 \left[A(0)-A(l)\right] = 2 \left[\sigma^2-A(l)\right]\,.
\end{equation}
Since the autocorrelation function vanishes on large scales close to our periodic box size ($l\to L$), the 2nd order structure function of a variable can be used to measure the standard deviation $\sigma$ of this variable because $\mathrm{SF}_2(l\to L) = 2 \sigma^2$. In our case, we obtained the standard deviations of the density $\sigma_\rho\!\sim\!1.88$ for solenoidal and $\sigma_\rho\!\sim\!5.95$ for compressive forcing simply by evaluating $\sigma_\rho\,=\,[0.5\,\mathrm{SF}_2(l\,=\,0.5\,L)]^{1/2}$ from Figure~\ref{fig:sf}. Note that for periodic boxes, the maximum length scale for measuring spatial correlations is half of the box size $L$. As for the power spectra, this is in good agreement with the independent analysis of the density PDFs \citep{FederrathKlessenSchmidt2008}.

The best-fit values of power-law exponents $\mathrm{SF}_2(l)\propto l^\eta$ of the structure functions in the inertial range are indicated in Figure~\ref{fig:sf} as well. Since $\mathrm{SF}_2(l)\propto l^\eta$ is the Fourier transform of the 1-dimensional Fourier spectrum $P(k)\propto k^{-\alpha}$, it follows that $\alpha=\eta+1$. For compressive forcing, the power-law scaling is in agreement with the $\Delta$-variance and Fourier spectra estimates. The corresponding value for the 3-dimensional density power spectrum derived from the structure function is $\beta=\alpha+2=\eta+3\!\sim\!3.47$ (see Table~1). For solenoidal forcing on the other hand, the best-fit value derived from the structure functions is $\sim\!3.24$ is in significant disagreement with the estimates from the Fourier spectra and $\Delta$-variance ($\beta\!\sim\!2.8$). \citet{StutzkiEtAl1998} provide the mathematical explanation for this. For a power-law scaling of Fourier spectrum functions with power-law exponent $\beta<E$ (here $E=3$), i.e. $\alpha<1$, the $E$-dimensional two-point correlation function (structure function) does not necessarily increase in a power-law fashion \citep[][Appendix~B]{StutzkiEtAl1998}. This limits the applicability of structure functions for estimating the inertial range scaling to density Fourier power-law exponents $E<\beta<E+2$.

\subsection{Mass Size Method}

\begin{figure}[t]
\centerline{\includegraphics[width=0.7\linewidth]{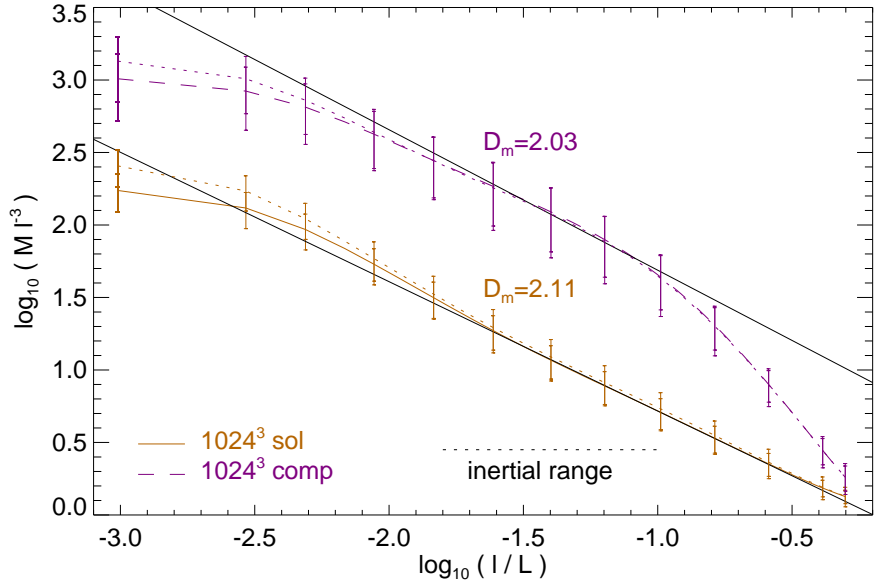}}
\caption{Log-log plot of the mass $M(l)$ compensated by $l^{-3}$ obtained by the mass size method described in Section~\ref{sec:mass-size} for solenoidal and compressive forcing. As in \citet{KowalLazarian2007}, a horizontal straight line therefore corresponds to a fractal mass dimension $D_m=3$. Power-law fits in the inertial range yield a fractal mass dimension $D_m\!\sim\!2.11$ for solenoidal forcing and $D_m\!\sim\!2.03$ for compressive forcing. However, the difference between solenoidal and compressive forcing inferred from the mass size method is not significant due to the large uncertainties (error bars). These large uncertainties are a result of the strong temporal fluctuations of the maximum density (Fig.~\ref{fig:timeevol}), since the mass size method relies on the density peaks, i.e., all cells with density $\rho>\rho_\mathrm{max}/2$ are used as a basis for computing the mass size relation $M(l)$. The dotted lines show $M\,l^{-3}$ using $\rho_\mathrm{max}$ only, i.e., without averaging over cells with $\rho>\rho_\mathrm{max}/2$. There is a systematic decrease of $M(l)$  with decreasing averaging threshold, which does not affect the inertial range scaling as long as the averaging is performed over all cells with $\rho>\rho_\mathrm{max}/2$.}
\label{fig:dm}
\end{figure}

Figure~\ref{fig:dm} shows the results obtained by applying the mass size method as described in Section~\ref{sec:mass-size}. In rough agreement with the results of the methods discussed so far, compressive forcing yields a smaller mass size exponent $D_m\!\sim\!2.03$ compared to solenoidal forcing with $D_m\!\sim\!2.11$ in the inertial range (Table~1). Unlike the other methods, however, this difference is not significant. The large 1$\sigma$ error is a direct consequence of the strong temporal fluctuations of the maximum density seen in Figure~\ref{fig:timeevol}. Since the mass size relation $M(l)$ is computed by growing concentric boxes centered on density peaks with $\rho>\rho_\mathrm{max}/2$, the mass is expected to fluctuate strongly, following the temporal fluctuations of the density peaks. Our results are therefore roughly consistent with the mass size analysis by \citet{KowalLazarian2007} and \citet{KritsukEtAl2007}.

\subsection{Box Counting Method}

\begin{figure}[t]
\centerline{\includegraphics[width=0.7\linewidth]{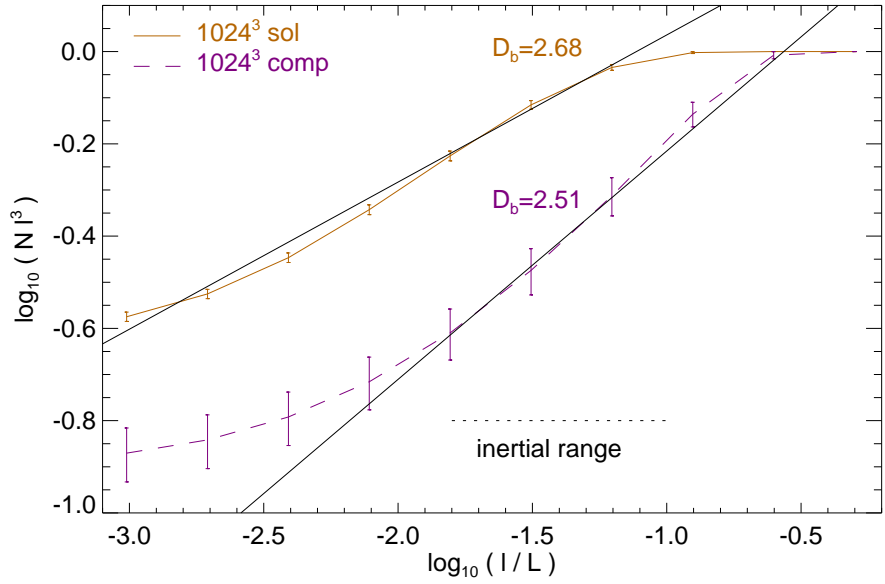}}
\caption{Log-log plot of $N(l)$ compensated with $l^3$ obtained by the box counting analysis described in Section~\ref{sec:box-counting} for solenoidal and compressive forcing respectively. A horizontal straight line would correspond to a box dimension of $D_b=3$. The power-law exponents $D_b$ obtained from fits within the inertial range are drawn as straight lines. Compressive forcing yields a significantly smaller box dimension $D_b\!\sim\!2.51$ compared to solenoidal forcing ($D_b\!\sim\!2.68$).}
\label{fig:db}
\end{figure}

\begin{figure*}[t]
\centerline{
\includegraphics[width=0.5\linewidth]{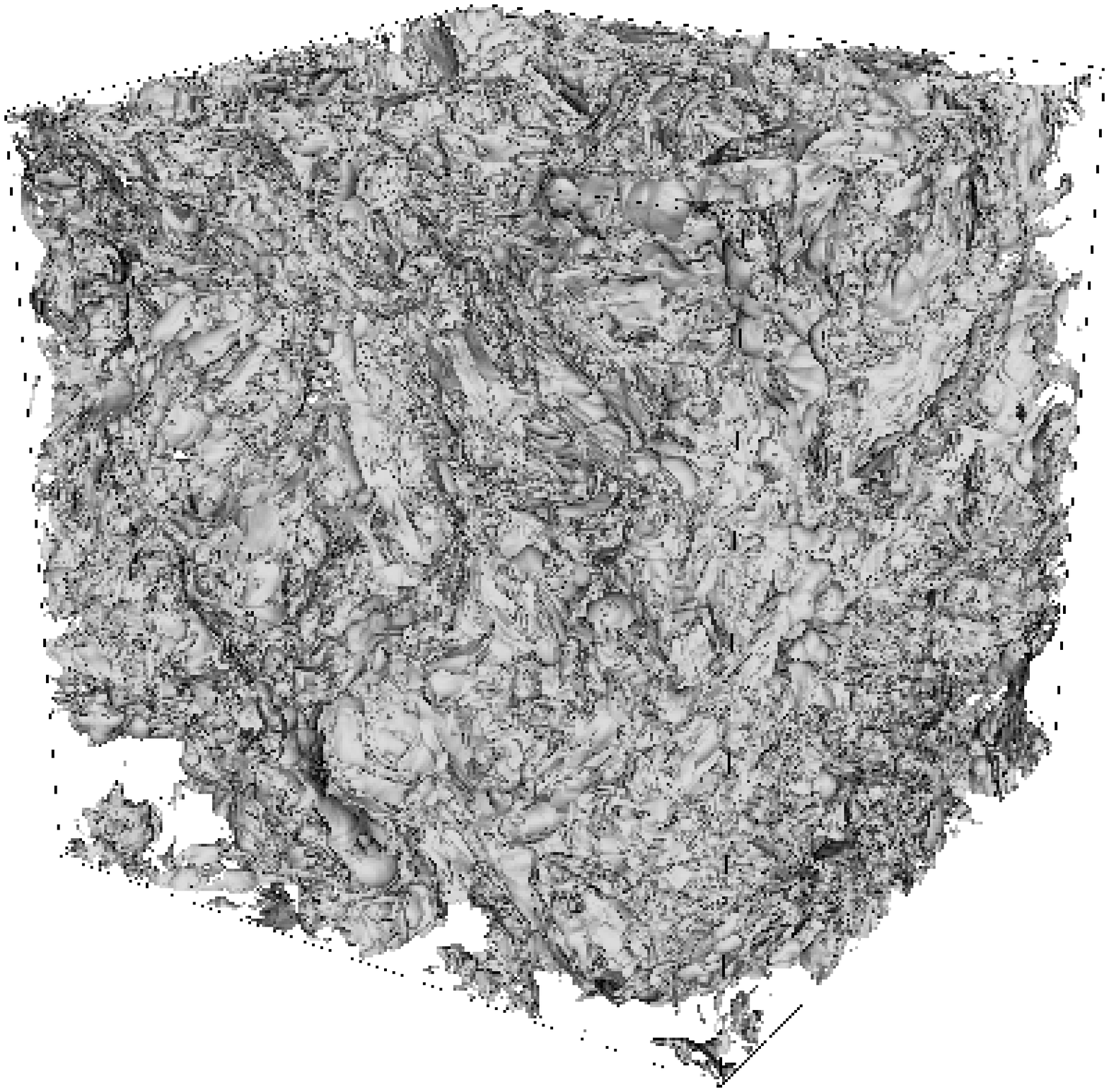}
\includegraphics[width=0.5\linewidth]{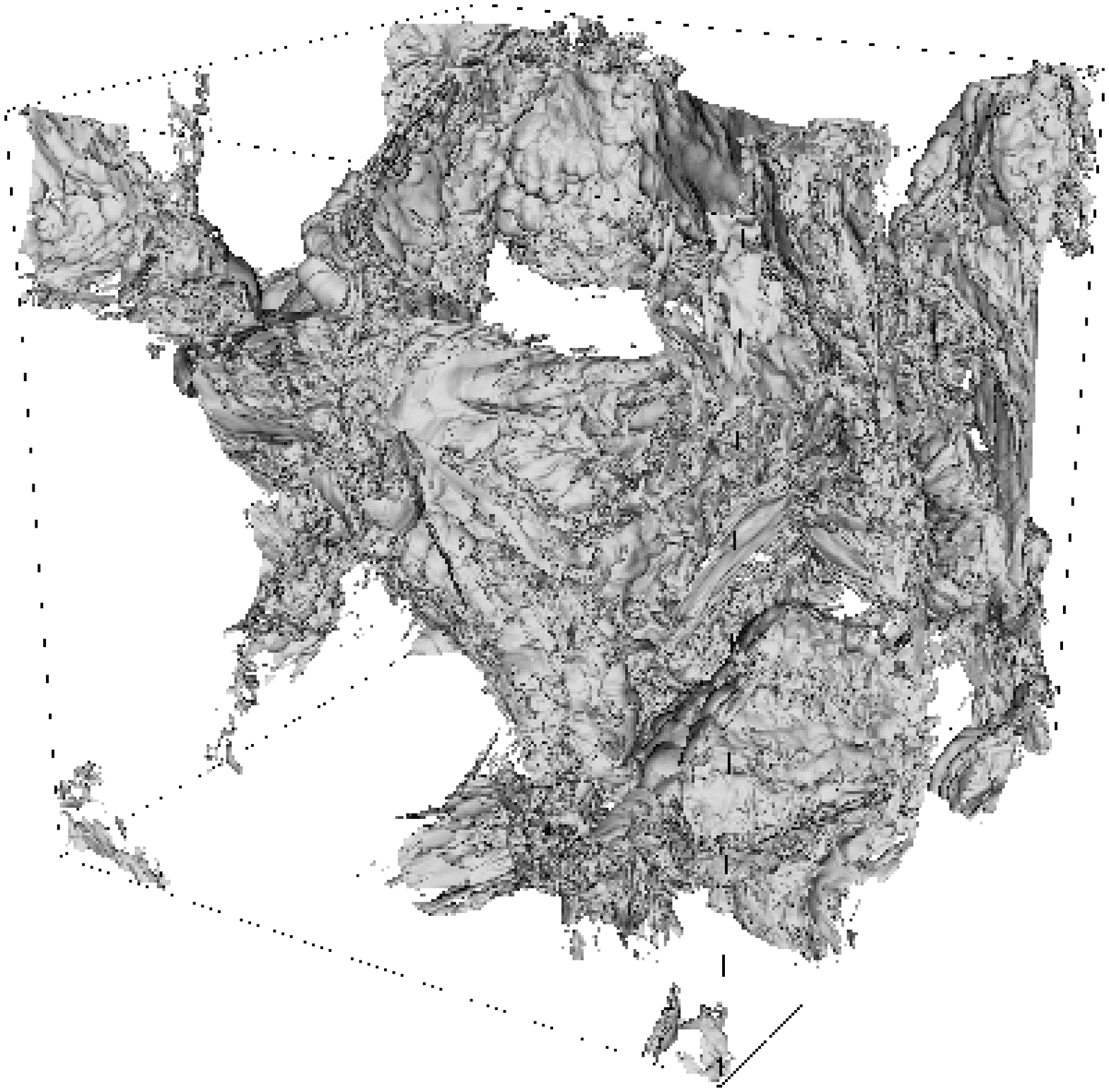}
}
\caption{Isosurface plots of the fractal structure defined by the mean density for solenoidal (\emph{left}) and compressive forcing (\emph{right}). The presence of hierarchical visual complexity indicates a fractal structure that can be analyzed with the box counting method \citep{MandelbrotFrame2002}.}
\label{fig:boxcountingimages}
\end{figure*}

\begin{figure*}[t]
\centerline{\includegraphics[width=1.0\linewidth]{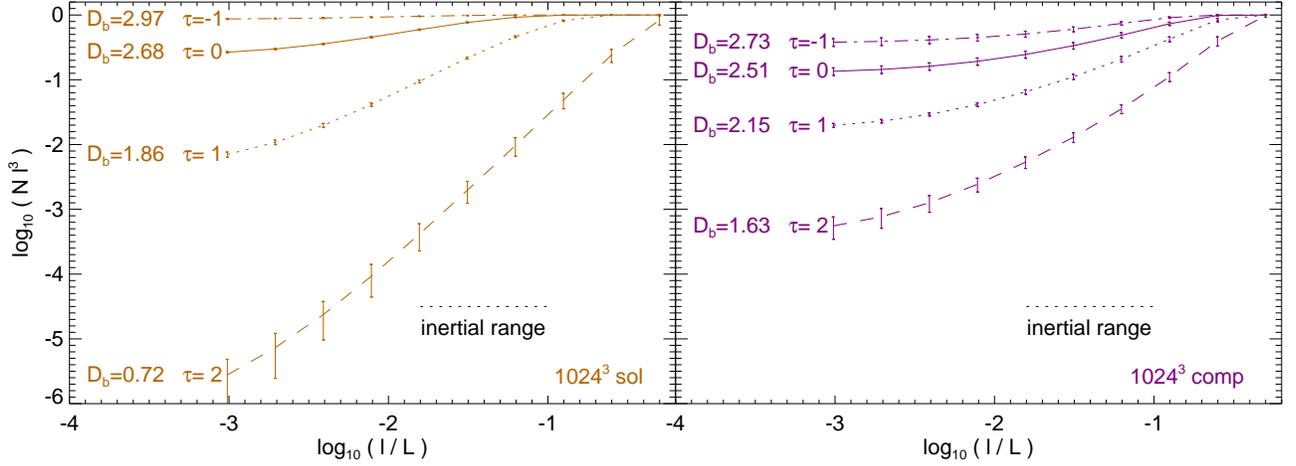}}
\caption{Shows the dependence of the box counting dimension on the threshold density $\rho_\mathrm{th}$ defining the fractal structure. Computing the fractal box counting dimension for $\tau\equiv\log_{10}(\rho_\mathrm{th}/\rho_0)=-1,\,0,\,1,\,2$ reveals the strong dependence of the inferred fractal dimension on the defining density threshold. \emph{Left panel:} solenoidal forcing; \emph{Right panel:} compressive forcing.}
\label{fig:dbthresh}
\end{figure*}

The results of the analysis using the box counting method as explained in Section~\ref{sec:box-counting} are presented in Figure~\ref{fig:db}. In this case, the fractal structure was defined by the mean density $\rho_0=1$ as threshold density for both solenoidal and compressive forcing. The analyzed structure as defined by $\rho_0$ is shown in Figure~\ref{fig:boxcountingimages}. We obtain the box counting dimensions $D_b\!\sim\!2.68$ for solenoidal, and $D_b\!\sim\!2.51$ for compressive forcing in the inertial range (Table~1). As discussed in Section~\ref{sec:box-counting}, the box counting dimension depends on the threshold for defining the fractal structure. We have checked its dependence on the threshold $\rho_\mathrm{th}$ by varying $\tau\equiv\log_{10}(\rho_\mathrm{th}/\rho_0)$. The results obtained by computing the box counting dimension for $\tau\,=\,-1,\,0,\,1,\,2$ are shown in Figure~\ref{fig:dbthresh} for solenoidal forcing (left panel) and compressive forcing (right panel). Note that $\tau=0$ corresponds to the mean density as defining threshold. As expected, the box dimension strongly depends on $\rho_\mathrm{th}$. Significant differences between solenoidal and compressive forcing are obtained for different threshold densities. For thresholds $\tau\gtrsim 0$, the box dimension is smaller for solenoidal forcing than for compressive forcing in contrast to $\tau\lesssim 0$. The latter is as a consequence of the much more space filling density structure for the solenoidal case (see Fig.~\ref{fig:snapshots}). On the other hand, for $\tau=2$ the solenoidal case yields filamentary structures with small fractal box dimension ($D_b\!\sim\!0.72$), while structures in the compressive case are coherent elongated and almost sheetlike structures with larger fractal dimension ($D_b\!\sim\!1.63$). Obtaining absolute estimates for the fractal dimension using the box counting method in the present context is consequently impossible. However, differences among different data sets, e.g., solenoidal versus compressive forcing can be measured with the box counting method, if the same defining density threshold is used.

\subsection{Perimeter Area Method}

In this section, we show results of the perimeter area method described in Section~\ref{sec:perimeter-area}. This method is often applied to observational data sets to infer the fractal dimension of interstellar clouds. Although we are aware of the fact that our simulations can only provide a very limited approximation to real interstellar gas, we nevertheless are convinced that comparison with observational data will provide physical insight. The perimeters of interstellar gas clouds exhibit fractal dimensions in the range $D_p\!\sim\!1.2\dots1.6$ \citep[e.g.,][]{Beech1987,BazellDesert1988,DickmanEtAl1990,FalgaroneEtAl1991,VogelaarWakkerSchwarz1991,HetemLepine1993,VogelaarWakkerAl1994,WestpfahlEtAl1999,KimEtAl2003,Lee2004,SanchezEtAl2007} with most of the studies finding $D_p\sim1.3\dots1.4$. These results are always obtained from projected images. A plausible conversion to the 3-dimensional fractal dimension, $D\!\sim\!D_p+1$ is discussed by \citet{Beech1992}. However, this relation holds rigorously only for the analysis of slices through an isotropic 3-dimensional monofractal, while interstellar clouds could be multifractals \citep[e.g.,][]{ChappellScalo2001}. As discussed by \citet{StutzkiEtAl1998} and shown by \citet{SanchezEtAl2005}, $D_p+1$ can be different from the 3-dimensional fractal dimension for projected images.

\begin{figure}[t]
\centerline{\includegraphics[width=0.7\linewidth]{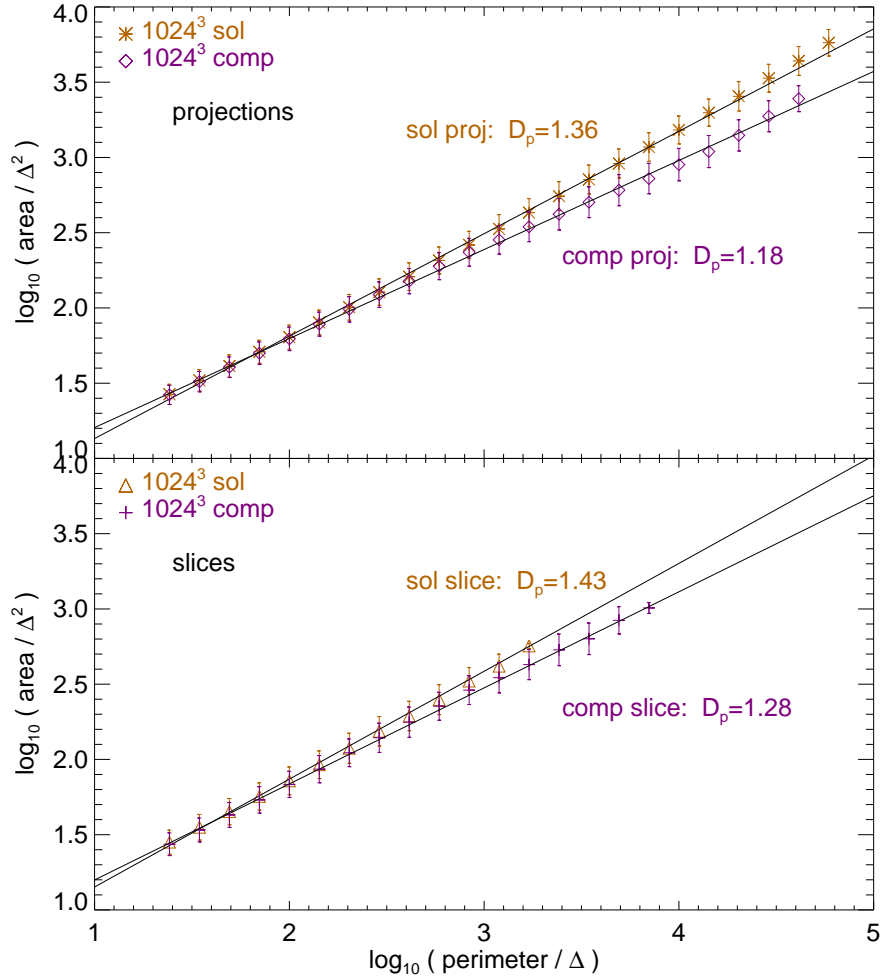}}
\caption{\emph{Top panel:} Perimeter area method applied to projections along $x$, $y$ and $z$ for solenoidal and compressive forcing respectively. \emph{Bottom panel:} Same as top panel but for slices at $x=0$, $y=0$ and $z=0$. The perimeter is given in units of the numerical cell size $\Delta=L/1024$.}
\label{fig:dp}
\end{figure}

We applied the perimeter area method to projections along the $x$-, $y$- and $z$-axis, as well as to slices at $x=0$, $y=0$ and $z=0$ of our simulation data cubes. The results are presented in Figure~\ref{fig:dp} for the projections (top panel) and the slices (bottom panel). Best-fit power laws to the projected data yield $D_p\!\sim\!1.36$ for solenoidal forcing and $D_p\!\sim\!1.18$ for compressive forcing, whereas for the slices we find $D_p\!\sim\!1.43$ and $D_p\!\sim\!1.28$, respectively. Thus, we find that the projections yield perimeter area dimensions systematically smaller than the slices for both forcings (Table~1). In agreement with the results obtained from the density Fourier spectra and the $\Delta$-variance analysis, the perimeter area method yields a significantly smaller perimeter area dimension for compressive forcing compared to solenoidal forcing.

\section{CONCLUSIONS} \label{sec:conclusions}

We have presented results of two high-resolution ($1024^3$ grid cells) hydrodynamic simulations of supersonic isothermal turbulence driven to rms Mach numbers $\mathcal{M}\!\sim\!5.5$. The first simulation uses the typically adopted solenoidal (divergence-free) forcing to excite turbulent motions, whereas the second one uses compressive (curl-free) forcing. We have shown that compressive forcing yields much stronger density contrasts compared to solenoidal forcing for the same rms Mach number. This implies that the turbulence production mechanism leaves a strong imprint on compressible turbulence statistics, especially altering the density statistics. Our results particularly suggests that the mixture of solenoidal and compressive modes of the turbulence forcing must be taken into account. We summarize our results as follows:

\begin{itemize}

\item The velocity Fourier spectra exhibit power laws in the inertial range for solenoidal and compressive forcing. The slopes obtained for both forcings are significantly steeper ($\sim\!1.9$) compared to the Kolmogorov slope ($5/3$), in agreement with previous studies \citep[e.g.,][]{KritsukEtAl2007,SchmidtEtAl2008} and in agreement with velocity dispersion to size relations inferred from observations \citep[e.g.,][]{Larson1981,FalgaronePugetPerault1992,HeyerBrunt2004,PadoanEtAl2006}.

\item From the integral of the density fluctuation Fourier spectra and from the asymptotic behavior of the 2nd order density structure function, we obtained the standard deviation of the density distribution $\sigma_\rho$. Compressive forcing yields a standard deviation $\sim\!$ three times larger compared to solenoidal forcing, in agreement with the results found in our previous study analyzing density probability distribution functions \citep{FederrathKlessenSchmidt2008} and in agreement with the studies by \citet{PassotVazquez1998}, \citet{KritsukEtAl2007}, \citet{BeetzEtAl2008} and \citet{SchmidtEtAl2008}.

\item The density fluctuation Fourier spectra are significantly steeper for compressive forcing in the inertial range compared to solenoidal forcing. Consistent results were obtained using complementary analysis methods, i.e., by comparing the $\Delta$-variances \citep{OssenkopfKripsStutzki2008a} and the 2nd order structure functions of the density field. Our estimates of density spectra for solenoidal forcing are in agreement with previous studies, e.g., the weakly magnetized super-Alfv\'enic supersonic MHD models by \citet{PadoanEtAl2004} and \citet{KowalLazarianBeresnyak2007}, and consistent with the hydrodynamic estimates by \citet{KritsukNormanPadoan2006}. Although a comparison with observational results must be regarded with caution due to systematic uncertainties, our results for solenoidal and compressive forcing are in the range of inferred scaling exponents by observations \citep[e.g.,][]{BenschStutzkiOssenkopf2001}.

\item From the scaling of the density fluctuation Fourier spectra and the $\Delta$-variance applied to the 3-dimensional data and applied to 2-dimensional projections, we obtained fractal Hurst exponents following the analysis by \citet{StutzkiEtAl1998}. This implies fractal box counting and fractal perimeter area dimensions significantly smaller for compressive forcing compared to solenoidal forcing (see Table~1).

\item We analyzed the density structure using the fractal mass size method as introduced by \citet{KritsukEtAl2007}. Compressive forcing yields a smaller fractal mass dimension compared to solenoidal forcing. The mass size method is, however, particularly sensitive to the temporal fluctuations of density peaks. Given the large uncertainties, our results using this method are roughly consistent with the estimates by \citet{KritsukEtAl2007} and \citet{KowalLazarian2007}.

\item We analyzed the fractal density structure using the box counting method described in Section~\ref{sec:box-counting} and the perimeter area method (Section~\ref{sec:perimeter-area}) applied to projections and slices. We recover the significant differences between solenoidal and compressive forcing inferred from the density spectra and $\Delta$-variance analysis. However, the box counting dimension varies strongly with the defining density threshold. The perimeter area dimensions obtained from slices are roughly consistent with the computed perimeter area dimensions from the $\Delta$-variance given the systematic uncertainties (of order $\sim\!0.1$ for fractal dimension estimates) comparing different methods. The range of fractal dimensions obtained is consistent with the observations analyzed by \citet{ElmegreenFalgarone1996} suggesting an overall fractal dimension of interstellar clouds in the range $D\!\sim\!2.3\pm0.3$.

\end{itemize}

\acknowledgements
We thank Volker Ossenkopf and Mordecai-Mark Mac Low for providing us with the $\Delta$-variance tool for analyzing 3-dimensional data sets. We are grateful to Nestor S{\'a}nchez for making available his tool for computing the perimeter area dimension in periodic data sets. We thank the anonymous referees for helpful comments, which improved the manuscript. CF acknowledges financial support by the International Max Planck Research School for Astronomy and Cosmic Physics (IMPRS-A) and the Heidelberg Graduate School of Fundamental Physics (HGSFP). The HGSFP is funded by the Excellence Initiative of the German Research Foundation DFG GSC 129/1. RSK thanks for support from the Emmy Noether grant KL 1358/1. CF and RSK acknowledge subsidies from the DFG SFB 439 Galaxies in the Early Universe. The simulations used resources from HLRBII project h0972 at Leibniz Rechenzentrum Garching. The software used in this work was in part developed by the DOE-supported ASC / Alliance Center for Astrophysical Thermonuclear Flashes at the University of Chicago.


\begin{thebibliography}{83}
\expandafter\ifx\csname natexlab\endcsname\relax\def\natexlab#1{#1}\fi

\bibitem[{{Ballesteros-Paredes} {et~al.}(2006){Ballesteros-Paredes}, {Gazol},
  {Kim}, {Klessen}, {Jappsen}, \& {Tejero}}]{BallesterosEtAl2006}
{Ballesteros-Paredes}, J., {Gazol}, A., {Kim}, J., {Klessen}, R.~S., {Jappsen},
  A.-K., \& {Tejero}, E. 2006, \apj, 637, 384

\bibitem[{{Bazell} \& {Desert}(1988)}]{BazellDesert1988}
{Bazell}, D., \& {Desert}, F.~X. 1988, \apj, 333, 353

\bibitem[{{Beech}(1987)}]{Beech1987}
{Beech}, M. 1987, \apss, 133, 193

\bibitem[{{Beech}(1992)}]{Beech1992}
---. 1992, \apss, 192, 103

\bibitem[{{Beetz} {et~al.}(2008){Beetz}, {Schwarz}, {Dreher}, \&
  {Grauer}}]{BeetzEtAl2008}
{Beetz}, C., {Schwarz}, C., {Dreher}, J., \& {Grauer}, R. 2008, Physics Letters
  A, 372, 3037

\bibitem[{{Bensch} {et~al.}(2001){Bensch}, {Stutzki}, \&
  {Ossenkopf}}]{BenschStutzkiOssenkopf2001}
{Bensch}, F., {Stutzki}, J., \& {Ossenkopf}, V. 2001, \aap, 366, 636

\bibitem[{{Boldyrev} {et~al.}(2002){Boldyrev}, {Nordlund}, \&
  {Padoan}}]{BoldyrevNordlundPadoan2002}
{Boldyrev}, S., {Nordlund}, {\AA}., \& {Padoan}, P. 2002, \apj, 573, 678

\bibitem[{{Chappell} \& {Scalo}(2001)}]{ChappellScalo2001}
{Chappell}, D., \& {Scalo}, J. 2001, \apj, 551, 712

\bibitem[{{Colella} \& {Woodward}(1984)}]{ColellaWoodward1984}
{Colella}, P., \& {Woodward}, P.~R. 1984, Journal of Computational Physics, 54,
  174

\bibitem[{{Dib} {et~al.}(2008){Dib}, {Brandenburg}, {Kim}, {Gopinathan}, \&
  {Andr{\'e}}}]{DibEtAl2008}
{Dib}, S., {Brandenburg}, A., {Kim}, J., {Gopinathan}, M., \& {Andr{\'e}}, P.
  2008, \apjl, 678, L105

\bibitem[{{Dickman} {et~al.}(1990){Dickman}, {Horvath}, \&
  {Margulis}}]{DickmanEtAl1990}
{Dickman}, R.~L., {Horvath}, M.~A., \& {Margulis}, M. 1990, \apj, 365, 586

\bibitem[{Dobler {et~al.}(2003)Dobler, Haugen, Yousef, \&
  Brandenburg}]{DoblerEtAl2003}
Dobler, W., Haugen, N. E.~L., Yousef, T.~A., \& Brandenburg, A. 2003, Phys.
  Rev. E, 68, 026304

\bibitem[{{Dubey} {et~al.}(2008){Dubey}, {Fisher}, {Graziani}, {Jordan},
  {Lamb}, {Reid}, {Rich}, {Sheeler}, {Townsley}, \& {Weide}}]{DubeyEtAl2008}
{Dubey}, A., {Fisher}, R., {Graziani}, C., {Jordan}, IV, G.~C., {Lamb}, D.~Q.,
  {Reid}, L.~B., {Rich}, P., {Sheeler}, D., {Townsley}, D., \& {Weide}, K.
  2008, in Astronomical Society of the Pacific Conference Series, Vol. 385,
  Numerical Modeling of Space Plasma Flows, ed. N.~V. {Pogorelov}, E.~{Audit},
  \& G.~P. {Zank}, 145--+

\bibitem[{{Dubinski} {et~al.}(1995){Dubinski}, {Narayan}, \&
  {Phillips}}]{DubinskiNarayanPhillips1995}
{Dubinski}, J., {Narayan}, R., \& {Phillips}, T.~G. 1995, \apj, 448, 226

\bibitem[{{Elmegreen} \& {Falgarone}(1996)}]{ElmegreenFalgarone1996}
{Elmegreen}, B.~G., \& {Falgarone}, E. 1996, \apj, 471, 816

\bibitem[{{Elmegreen} \& {Scalo}(2004)}]{ElmegreenScalo2004}
{Elmegreen}, B.~G., \& {Scalo}, J. 2004, \araa, 42, 211

\bibitem[{{Esquivel} \& {Lazarian}(2005)}]{EsquivelLazarian2005}
{Esquivel}, A., \& {Lazarian}, A. 2005, \apj, 631, 320

\bibitem[{{Eswaran} \& {Pope}(1988)}]{EswaranPope1988}
{Eswaran}, V., \& {Pope}, S.~B. 1988, Computers and Fluids, 16, 257

\bibitem[{{Falgarone} {et~al.}(1994){Falgarone}, {Lis}, {Phillips}, {Pouquet},
  {Porter}, \& {Woodward}}]{FalgaroneEtAl1994}
{Falgarone}, E., {Lis}, D.~C., {Phillips}, T.~G., {Pouquet}, A., {Porter},
  D.~H., \& {Woodward}, P.~R. 1994, \apj, 436, 728

\bibitem[{{Falgarone} {et~al.}(1991){Falgarone}, {Phillips}, \&
  {Walker}}]{FalgaroneEtAl1991}
{Falgarone}, E., {Phillips}, T.~G., \& {Walker}, C.~K. 1991, \apj, 378, 186

\bibitem[{{Falgarone} {et~al.}(1992){Falgarone}, {Puget}, \&
  {Perault}}]{FalgaronePugetPerault1992}
{Falgarone}, E., {Puget}, J.-L., \& {Perault}, M. 1992, \aap, 257, 715

\bibitem[{{Federrath} {et~al.}(2008){Federrath}, {Klessen}, \&
  {Schmidt}}]{FederrathKlessenSchmidt2008}
{Federrath}, C., {Klessen}, R.~S., \& {Schmidt}, W. 2008, \apjl, 688, L79

\bibitem[{{Ferri{\`e}re}(2001)}]{Ferriere2001}
{Ferri{\`e}re}, K.~M. 2001, Reviews of Modern Physics, 73, 1031

\bibitem[{{Fleck}(1996)}]{Fleck1996}
{Fleck}, Jr., R.~C. 1996, \apj, 458, 739

\bibitem[{Frisch(1995)}]{Frisch1995}
Frisch, U. 1995, Turbulence (Cambridge: Cambridge Univ. Press)

\bibitem[{{Fryxell} {et~al.}(2000){Fryxell}, {Olson}, {Ricker}, {Timmes},
  {Zingale}, {Lamb}, {MacNeice}, {Rosner}, {Truran}, \&
  {Tufo}}]{FryxellEtAl2000}
{Fryxell}, B., {Olson}, K., {Ricker}, P., {Timmes}, F.~X., {Zingale}, M.,
  {Lamb}, D.~Q., {MacNeice}, P., {Rosner}, R., {Truran}, J.~W., \& {Tufo}, H.
  2000, \apjs, 131, 273

\bibitem[{{Haugen} \& {Brandenburg}(2004)}]{HaugenBrandenburg2004}
{Haugen}, N.~E., \& {Brandenburg}, A. 2004, \pre, 70, 026405

\bibitem[{{Heitsch} {et~al.}(2001){Heitsch}, {Mac Low}, \&
  {Klessen}}]{HeitschMacLowKlessen2001}
{Heitsch}, F., {Mac Low}, M.-M., \& {Klessen}, R.~S. 2001, \apj, 547, 280

\bibitem[{{Hetem} \& {Lepine}(1993)}]{HetemLepine1993}
{Hetem}, Jr., A., \& {Lepine}, J.~R.~D. 1993, \aap, 270, 451

\bibitem[{{Heyer} \& {Brunt}(2004)}]{HeyerBrunt2004}
{Heyer}, M.~H., \& {Brunt}, C.~M. 2004, \apjl, 615, L45

\bibitem[{{Hily-Blant} {et~al.}(2008){Hily-Blant}, {Falgarone}, \&
  {Pety}}]{HilyBlantFalgaronePety2008}
{Hily-Blant}, P., {Falgarone}, E., \& {Pety}, J. 2008, \aap, 481, 367

\bibitem[{{Jappsen} {et~al.}(2005){Jappsen}, {Klessen}, {Larson}, {Li}, \& {Mac
  Low}}]{JappsenEtAl2005}
{Jappsen}, A.-K., {Klessen}, R.~S., {Larson}, R.~B., {Li}, Y., \& {Mac Low},
  M.-M. 2005, \aap, 435, 611

\bibitem[{{Kim} {et~al.}(2003){Kim}, {Staveley-Smith}, {Dopita}, {Sault},
  {Freeman}, {Lee}, \& {Chu}}]{KimEtAl2003}
{Kim}, S., {Staveley-Smith}, L., {Dopita}, M.~A., {Sault}, R.~J., {Freeman},
  K.~C., {Lee}, Y., \& {Chu}, Y.-H. 2003, \apjs, 148, 473

\bibitem[{{Klein} {et~al.}(2007){Klein}, {Inutsuka}, {Padoan}, \&
  {Tomisaka}}]{KleinEtAl2007}
{Klein}, R.~I., {Inutsuka}, S.-I., {Padoan}, P., \& {Tomisaka}, K. 2007, in
  Protostars and Planets V, ed. B.~{Reipurth}, D.~{Jewitt}, \& K.~{Keil},
  99--116

\bibitem[{{Klessen}(2000)}]{Klessen2000}
{Klessen}, R.~S. 2000, \apj, 535, 869

\bibitem[{{Klessen} {et~al.}(2000){Klessen}, {Heitsch}, \& {Mac
  Low}}]{KlessenHeitschMacLow2000}
{Klessen}, R.~S., {Heitsch}, F., \& {Mac Low}, M.-M. 2000, \apj, 535, 887

\bibitem[{{Kolmogorov}(1941)}]{Kolmogorov1941c}
{Kolmogorov}, A.~N. 1941, Dokl. Akad. Nauk SSSR, 32, 16

\bibitem[{{Kowal} \& {Lazarian}(2007)}]{KowalLazarian2007}
{Kowal}, G., \& {Lazarian}, A. 2007, \apjl, 666, L69

\bibitem[{{Kowal} {et~al.}(2007){Kowal}, {Lazarian}, \&
  {Beresnyak}}]{KowalLazarianBeresnyak2007}
{Kowal}, G., {Lazarian}, A., \& {Beresnyak}, A. 2007, \apj, 658, 423

\bibitem[{{Kritsuk} {et~al.}(2006){Kritsuk}, {Norman}, \&
  {Padoan}}]{KritsukNormanPadoan2006}
{Kritsuk}, A.~G., {Norman}, M.~L., \& {Padoan}, P. 2006, \apjl, 638, L25

\bibitem[{{Kritsuk} {et~al.}(2007){Kritsuk}, {Norman}, {Padoan}, \&
  {Wagner}}]{KritsukEtAl2007}
{Kritsuk}, A.~G., {Norman}, M.~L., {Padoan}, P., \& {Wagner}, R. 2007, \apj,
  665, 416

\bibitem[{{Larson}(1981)}]{Larson1981}
{Larson}, R.~B. 1981, \mnras, 194, 809

\bibitem[{{Lee}(2004)}]{Lee2004}
{Lee}, Y. 2004, Journal of Korean Astronomical Society, 37, 137

\bibitem[{{Lemaster} \& {Stone}(2008)}]{LemasterStone2008}
{Lemaster}, M.~N., \& {Stone}, J.~M. 2008, \apjl, 682, L97

\bibitem[{{Li} {et~al.}(2003){Li}, {Klessen}, \& {Mac
  Low}}]{LiKlessenMacLow2003}
{Li}, Y., {Klessen}, R.~S., \& {Mac Low}, M.-M. 2003, \apj, 592, 975

\bibitem[{{Mac Low}(1999)}]{MacLow1999}
{Mac Low}, M.-M. 1999, \apj, 524, 169

\bibitem[{{Mac Low} \& {Klessen}(2004)}]{MacLowKlessen2004}
{Mac Low}, M.-M., \& {Klessen}, R.~S. 2004, Reviews of Modern Physics, 76, 125

\bibitem[{{Mac Low} {et~al.}(1998){Mac Low}, {Klessen}, {Burkert}, \&
  {Smith}}]{MacLowEtAl1998}
{Mac Low}, M.-M., {Klessen}, R.~S., {Burkert}, A., \& {Smith}, M.~D. 1998,
  Physical Review Letters, 80, 2754

\bibitem[{{Mac Low} \& {Ossenkopf}(2000)}]{MacLowOssenkopf2000}
{Mac Low}, M.-M., \& {Ossenkopf}, V. 2000, \aap, 353, 339

\bibitem[{{Mandelbrot}(1983)}]{Mandelbrot1983}
{Mandelbrot}, B.~B. 1983, {The fractal geometry of nature} (New York: Freeman)

\bibitem[{{Mandelbrot} \& {Frame}(2002)}]{MandelbrotFrame2002}
{Mandelbrot}, B.~B., \& {Frame}, M. 2002, {Fractals} (San Diego, Calif.:
  Encyclopedia of Physical Science and Technology, Academic Press)

\bibitem[{{Myers}(1983)}]{Myers1983}
{Myers}, P.~C. 1983, \apj, 270, 105

\bibitem[{{Offner} {et~al.}(2008){Offner}, {Klein}, \&
  {McKee}}]{OffnerKleinMcKee2008}
{Offner}, S.~S.~R., {Klein}, R.~I., \& {McKee}, C.~F. 2008, \apj, 686, 1174

\bibitem[{{Ossenkopf} {et~al.}(2006){Ossenkopf}, {Esquivel}, {Lazarian}, \&
  {Stutzki}}]{OssenkopfEtAl2006}
{Ossenkopf}, V., {Esquivel}, A., {Lazarian}, A., \& {Stutzki}, J. 2006, \aap,
  452, 223

\bibitem[{{Ossenkopf} {et~al.}(2001){Ossenkopf}, {Klessen}, \&
  {Heitsch}}]{OssenkopfKlessenHeitsch2001}
{Ossenkopf}, V., {Klessen}, R.~S., \& {Heitsch}, F. 2001, \aap, 379, 1005

\bibitem[{{Ossenkopf} {et~al.}(2008){Ossenkopf}, {Krips}, \&
  {Stutzki}}]{OssenkopfKripsStutzki2008a}
{Ossenkopf}, V., {Krips}, M., \& {Stutzki}, J. 2008, \aap, 485, 917

\bibitem[{{Ossenkopf} \& {Mac Low}(2002)}]{OssenkopfMacLow2002}
{Ossenkopf}, V., \& {Mac Low}, M.-M. 2002, \aap, 390, 307

\bibitem[{{Ostriker} {et~al.}(2001){Ostriker}, {Stone}, \&
  {Gammie}}]{OstrikerStoneGammie2001}
{Ostriker}, E.~C., {Stone}, J.~M., \& {Gammie}, C.~F. 2001, \apj, 546, 980

\bibitem[{{Padoan} {et~al.}(2003){Padoan}, {Boldyrev}, {Langer}, \&
  {Nordlund}}]{PadoanEtAl2003}
{Padoan}, P., {Boldyrev}, S., {Langer}, W., \& {Nordlund}, {\AA}. 2003, \apj,
  583, 308

\bibitem[{{Padoan} {et~al.}(2004{\natexlab{a}}){Padoan}, {Jimenez}, {Juvela},
  \& {Nordlund}}]{PadoanEtAl2004}
{Padoan}, P., {Jimenez}, R., {Juvela}, M., \& {Nordlund}, {\AA}.
  2004{\natexlab{a}}, \apjl, 604, L49

\bibitem[{{Padoan} {et~al.}(2004{\natexlab{b}}){Padoan}, {Jimenez}, {Nordlund},
  \& {Boldyrev}}]{PadoanJimenezNordlundBoldyrev2004}
{Padoan}, P., {Jimenez}, R., {Nordlund}, {\AA}., \& {Boldyrev}, S.
  2004{\natexlab{b}}, Physical Review Letters, 92, 191102

\bibitem[{{Padoan} {et~al.}(2006){Padoan}, {Juvela}, {Kritsuk}, \&
  {Norman}}]{PadoanEtAl2006}
{Padoan}, P., {Juvela}, M., {Kritsuk}, A., \& {Norman}, M.~L. 2006, \apjl, 653,
  L125

\bibitem[{{Padoan} {et~al.}(1997){Padoan}, {Nordlund}, \&
  {Jones}}]{PadoanNordlundJones1997}
{Padoan}, P., {Nordlund}, {\AA}., \& {Jones}, B.~J.~T. 1997, \mnras, 288, 145

\bibitem[{{Passot} {et~al.}(1988){Passot}, {Pouquet}, \&
  {Woodward}}]{PassotPouquetWoodward1988}
{Passot}, T., {Pouquet}, A., \& {Woodward}, P. 1988, \aap, 197, 228

\bibitem[{{Passot} \& {V{\'a}zquez-Semadeni}(1998)}]{PassotVazquez1998}
{Passot}, T., \& {V{\'a}zquez-Semadeni}, E. 1998, \pre, 58, 4501

\bibitem[{{Pavlovski} {et~al.}(2006){Pavlovski}, {Smith}, \& {Mac
  Low}}]{PavlovskiSmithMacLow2006}
{Pavlovski}, G., {Smith}, M.~D., \& {Mac Low}, M.-M. 2006, \mnras, 368, 943

\bibitem[{Peitgen {et~al.}(2004)Peitgen, J\"urgens, \& Saupe}]{PeitgenEtAl2004}
Peitgen, H.-O., J\"urgens, H., \& Saupe, D. 2004, {Chaos and fractals - new
  frontiers of science} (New York; Berlin; Heidelberg: Springer)

\bibitem[{{Perault} {et~al.}(1986){Perault}, {Falgarone}, \&
  {Puget}}]{PeraultFalgaronePuget1986}
{Perault}, M., {Falgarone}, E., \& {Puget}, J.~L. 1986, \aap, 157, 139

\bibitem[{{Porter} {et~al.}(1992){Porter}, {Pouquet}, \&
  {Woodward}}]{PorterPouquetWoodward1992}
{Porter}, D.~H., {Pouquet}, A., \& {Woodward}, P.~R. 1992, Physical Review
  Letters, 68, 3156

\bibitem[{{S{\'a}nchez} {et~al.}(2005){S{\'a}nchez}, {Alfaro}, \&
  {P{\'e}rez}}]{SanchezEtAl2005}
{S{\'a}nchez}, N., {Alfaro}, E.~J., \& {P{\'e}rez}, E. 2005, \apj, 625, 849

\bibitem[{{S{\'a}nchez} {et~al.}(2007){S{\'a}nchez}, {Alfaro}, \&
  {P{\'e}rez}}]{SanchezEtAl2007}
---. 2007, \apj, 656, 222

\bibitem[{{Scalo} \& {Elmegreen}(2004)}]{ScaloElmegreen2004}
{Scalo}, J., \& {Elmegreen}, B.~G. 2004, \araa, 42, 275

\bibitem[{{Schmidt} {et~al.}(2009){Schmidt}, {Federrath}, {Hupp}, {Kern}, \&
  {Niemeyer}}]{SchmidtEtAl2008}
{Schmidt}, W., {Federrath}, C., {Hupp}, M., {Kern}, S., \& {Niemeyer}, J.~C.
  2009, \aap, 494, 127

\bibitem[{{Schmidt} {et~al.}(2006){Schmidt}, {Hillebrandt}, \&
  {Niemeyer}}]{SchmidtEtAl2006}
{Schmidt}, W., {Hillebrandt}, W., \& {Niemeyer}, J.~C. 2006, Computers and
  Fluids, 35, 353

\bibitem[{{Solomon} {et~al.}(1987){Solomon}, {Rivolo}, {Barrett}, \&
  {Yahil}}]{SolomonEtAl1987}
{Solomon}, P.~M., {Rivolo}, A.~R., {Barrett}, J., \& {Yahil}, A. 1987, \apj,
  319, 730

\bibitem[{{Stone} {et~al.}(1998){Stone}, {Ostriker}, \&
  {Gammie}}]{StoneOstrikerGammie1998}
{Stone}, J.~M., {Ostriker}, E.~C., \& {Gammie}, C.~F. 1998, \apjl, 508, L99

\bibitem[{{Stutzki} {et~al.}(1998){Stutzki}, {Bensch}, {Heithausen},
  {Ossenkopf}, \& {Zielinsky}}]{StutzkiEtAl1998}
{Stutzki}, J., {Bensch}, F., {Heithausen}, A., {Ossenkopf}, V., \& {Zielinsky},
  M. 1998, \aap, 336, 697

\bibitem[{{V{\'a}zquez-Semadeni}(1994)}]{Vazquez1994}
{V{\'a}zquez-Semadeni}, E. 1994, \apj, 423, 681

\bibitem[{{V{\'a}zquez-Semadeni} {et~al.}(2003){V{\'a}zquez-Semadeni},
  {Ballesteros-Paredes}, \& {Klessen}}]{VazquezBallesterosKlessen2003}
{V{\'a}zquez-Semadeni}, E., {Ballesteros-Paredes}, J., \& {Klessen}, R.~S.
  2003, \apjl, 585, L131

\bibitem[{{Vogelaar} \& {Wakker}(1994)}]{VogelaarWakkerAl1994}
{Vogelaar}, M.~G.~R., \& {Wakker}, B.~P. 1994, \aap, 291, 557

\bibitem[{{Vogelaar} {et~al.}(1991){Vogelaar}, {Wakker}, \&
  {Schwarz}}]{VogelaarWakkerSchwarz1991}
{Vogelaar}, M.~G.~R., {Wakker}, B.~P., \& {Schwarz}, U.~J. 1991, in IAU
  Symposium, Vol. 147, Fragmentation of Molecular Clouds and Star Formation,
  ed. E.~{Falgarone}, F.~{Boulanger}, \& G.~{Duvert}, 508--+

\bibitem[{{Westpfahl} {et~al.}(1999){Westpfahl}, {Coleman}, {Alexander}, \&
  {Tongue}}]{WestpfahlEtAl1999}
{Westpfahl}, D.~J., {Coleman}, P.~H., {Alexander}, J., \& {Tongue}, T. 1999,
  \aj, 117, 868

\bibitem[{{Wolfire} {et~al.}(1995){Wolfire}, {Hollenbach}, {McKee}, {Tielens},
  \& {Bakes}}]{WolfireEtAl1995}
{Wolfire}, M.~G., {Hollenbach}, D., {McKee}, C.~F., {Tielens}, A.~G.~G.~M., \&
  {Bakes}, E.~L.~O. 1995, \apj, 443, 152

\end{thebibliography}

\begin{table*}
\begin{center} \label{tab:results}
\caption{Power-law exponents and fractal dimension estimates comparing solenoidal and compressive forcing}
\def\arraystretch{1.1}
{\scriptsize 
\begin{tabular}{lccrr}
\hline
\hline
\noalign{\smallskip}
\hspace{4.5cm} & symbol & relation & solenoidal forcing & compressive forcing \\
\noalign{\smallskip} \hline \noalign{\smallskip}
1D power-law index for                      \\
velocity Fourier spectra~\mydotfill         & {$\epsilon$} & {$E_\mathrm{1D}(k)\propto k^{-\epsilon}$} & {$1.86^{\pm0.05}$} & {$1.94^{\pm0.05}$} \\
\noalign{\smallskip} \hline \noalign{\smallskip}
1D power-law index for density              \\
fluctuation Fourier spectra~\mydotfill      & {$\alpha$} & {$P_\mathrm{1D}(k)\propto k^{-\alpha}$} & {$0.78^{\pm0.06}$} & {$1.44^{\pm0.23}$} \\
\noalign{\smallskip} \hline \noalign{\smallskip}
3D power-law index for density              \\
fluctuation Fourier spectra~\mydotfill      & {$\beta$} & {$P_\mathrm{3D}(k)\propto k^{-\beta}\propto k^{-(\alpha+2)}$} & {$2.78^{\pm0.06}$} & {$3.44^{\pm0.23}$} \\
\noalign{\medskip}
derived Hurst exponent~\mydotfill           & $H$   & $H=(\beta-2)/2$ & $0.39^{\pm0.03}$ & $0.72^{\pm0.12}$ \\
box counting dimension~\mydotfill           & $D_b$ & $D_b=3-H$       & $2.61^{\pm0.03}$ & $2.28^{\pm0.12}$ \\
perimeter area dimension~\mydotfill         & $D_p$ & $D_p=2-H$       & $1.61^{\pm0.03}$ & $1.28^{\pm0.12}$ \\
\noalign{\smallskip} \hline \noalign{\smallskip}
power-law index for $\Delta$-variance       \\
applied to 3D data~\mydotfill               & {$\tilde{\beta}$} & {$\sigma_\Delta^2(l)\propto l^{\tilde{\beta}-2}$, $\tilde{\beta}\approx\beta$} & {$2.89^{\pm0.05}$} & {$3.44^{\pm0.19}$} \\
\noalign{\medskip}
derived Hurst exponent~\mydotfill           & $H$   & $H=(\tilde{\beta}-2)/2$ & $0.45^{\pm0.03}$ & $0.72^{\pm0.10}$ \\
box counting dimension~\mydotfill           & $D_b$ & $D_b=3-H$               & $2.55^{\pm0.03}$ & $2.28^{\pm0.10}$ \\
perimeter area dimension~\mydotfill         & $D_p$ & $D_p=2-H$               & $1.55^{\pm0.03}$ & $1.28^{\pm0.10}$ \\
\noalign{\smallskip} \hline \noalign{\smallskip}
power-law index for $\Delta$-variance       \\
applied to 2D projections~\mydotfill        & {$\hat{\beta}$} & {$\sigma_\Delta^2(l)\propto l^{\hat{\beta}-1}$, $\hat{\beta}\approx\beta$} & {$2.81^{\pm0.07}$} & {$3.37^{\pm0.21}$} \\
\noalign{\medskip}
derived Hurst exponent~\mydotfill           & $H$   & $H=(\hat{\beta}-2)/2$   & $0.41^{\pm0.04}$ & $0.69^{\pm0.11}$ \\
box counting dimension~\mydotfill           & $D_b$ & $D_b=3-H$               & $2.59^{\pm0.04}$ & $2.31^{\pm0.11}$ \\
perimeter area dimension~\mydotfill         & $D_p$ & $D_p=2-H$               & $1.59^{\pm0.04}$ & $1.31^{\pm0.11}$ \\
\noalign{\smallskip} \hline \noalign{\smallskip}
power-law index of 2nd order                \\
density structure function~\mydotfill       & {$\eta$} & {$\mathrm{SF}_2(l)\propto l^\eta \propto l^{\alpha-1}$, for $1<\alpha<3$} & {$0.24^{\pm0.03}$} & {$0.47^{\pm0.09}$} \\
\noalign{\smallskip} \hline \noalign{\smallskip}
mass size method averaged over              \\
density peaks with $\rho>\rho_\mathrm{max}/2$~\mydotfill & {$D_m$} & {$M(l)\propto l^{D_m}$} & {$2.11^{\pm0.16}$} & {$2.03^{\pm0.26}$} \\
\noalign{\smallskip} \hline \noalign{\smallskip}
box counting dimension with                 \\
$\rho_0$ as defining threshold~\mydotfill   & {$D_b$} & {$N(l)\propto l^{-D_b}$} & {$2.68^{\pm0.04}$} & {$2.51^{\pm0.08}$} \\
\noalign{\smallskip} \hline \noalign{\smallskip}
box counting dimension with                 \\
$\sigma_\rho$ as defining threshold~\mydotfill & {$\tilde{D}_b$} & {$N(l)\propto l^{-\tilde{D}_b}$} & {$2.51^{\pm0.05}$} & {$2.32^{\pm0.10}$} \\
\noalign{\smallskip} \hline \noalign{\smallskip}
perimeter area dimension                    \\
for 2D projections~\mydotfill               & {$D_p$} & {$\mathcal{P(A)}\propto\mathcal{A}^{D_p/2}$} & {$1.36^{\pm0.09}$} & {$1.18^{\pm0.10}$} \\
\noalign{\smallskip} \hline \noalign{\smallskip}
perimeter area dimension                    \\
for 2D slices~\mydotfill                    & {$\tilde{D}_p$} & {$\mathcal{P(A)}\propto\mathcal{A}^{\tilde{D}_p/2}$} & {$1.43^{\pm0.09}$} & {$1.28^{\pm0.11}$} \\
\noalign{\smallskip}
\hline
\hline
\end{tabular}
}
\end{center}
\end{table*}

\end{document}